\documentclass[12pt,a4paper]{article}
\usepackage[dvips]{graphicx}
\usepackage{geometry}
\usepackage{setspace}
\usepackage{textcomp}
\usepackage{amsfonts}
\usepackage{amsmath}
\usepackage{bm}
\usepackage{amsthm}
\usepackage{bigstrut}
\usepackage{multirow}
\usepackage{longtable}
\usepackage{booktabs}
\usepackage{xcolor}
\usepackage{comment}
\usepackage{hyperref}\hypersetup{colorlinks,allcolors=blue}
\usepackage[textsize=scriptsize]{todonotes} 
\usepackage{caption} 
\usepackage{subcaption}
\usepackage{float}
\newcommand*\diff{\mathop{}\!\mathrm{d}}

\DeclareGraphicsExtensions{.eps,.ps}

\begin{document}
\begin{titlepage}
\linespread{1.2}
\title{A Semi-Parametric Bayesian Generalized Least Squares Estimator}
\author{Ruochen Wu\thanks{Ruochen (first name) Wu (family name), School of Economics, Fudan University, Office 206, Building 11, 220 Handan Road, Fudan University, Shanghai 200433, China. Email: wuruochen@fudan.edu.cn}
\smallskip \\
\and Melvyn Weeks\thanks{Corresponding Author: Melvyn (first name) Weeks (family name), Faculty of Economics and Clare College, University of Cambridge, Cambridge CB3 9DD, UK. Email: mw217@econ.cam.ac.uk.}}
\date{}
\maketitle
\thispagestyle{empty}

\begin{abstract}
{\small
In this paper we propose a semi-parametric Bayesian Generalized Least Squares estimator. In a generic setting where each error is a vector, the parametric Generalized Least Square estimator maintains the assumption that each error vector has the same distributional parameters. In reality, however, errors are likely to be heterogeneous regarding their distributions. To cope with such heterogeneity, a Dirichlet process prior is introduced for the distributional parameters of the errors, leading to the error distribution being a mixture of a variable number of normal distributions. Our method let the number of normal components be data driven. Semi-parametric Bayesian estimators for two specific cases are then presented: the Seemingly Unrelated Regression for equation systems and the Random Effects Model for panel data. We design a series of simulation experiments to explore the performance of our estimators. The results demonstrate that our estimators obtain smaller posterior standard deviations and mean squared errors than the Bayesian estimators using a parametric mixture of normal distributions or a normal distribution. We then apply our semi-parametric Bayesian estimators for equation systems and panel data models to empirical data.

\medskip


\medskip

}

\end{abstract}
\end{titlepage}


\pagebreak

\section{Introduction}
The Generalized Least Square (\textsc{gls}) estimator is a family of econometric methods that have seen numerous applications in empirical economics. As pointed out by \hyperlink{Wooldridge, 2003}{Wooldridge (2003)}, parametric \textsc{gls} type estimators accommodate a deviation from the assumption that the errors in the model are homoskedastic and have no serial correlation. For example, compared to the ordinary least squares regression model, \textsc{gls} no  longer assumes that the covariance matrix of the errors is diagonal with identical diagonal elements.

In a more general setting, each error may be a random vector, which includes some of the most popular applications of \textsc{gls}. For example, the Seemingly Unrelated Regression (\textsc{sur}, \hyperlink{Zellner, 1962}{Zellner, 1962} and \hyperlink{Zellner, 1971}{1971}) has been developed for equation systems, and widely applied to gain efficiency by exploiting the correlation between errors across equations. Similarly, in the analysis of panel data the random effects model (\textsc{rem}) recognizes that there are individual specific, time-invariant features that are unobservable and uncorrelated with the explanatory variables.

However, the parametric \textsc{gls} still maintains the assumption that the error vector of each individual has the same covariance matrix. In reality, however, heterogeneity in error distributions is a major concern in empirical analyses. Such heterogeneity can be caused by observations on individuals or households reflecting variation in features such as the size of the household and the level of income, among others. It is a challenge for analysts who seek efficient estimates and inference with the data to capture the form of the heterogeneity in observations.
 
The standard Bayesian approach to \textsc{gls} assumes that the error distribution is multivariate normal. Recent developments in Bayesian methods allow the use of prior information to relax this assumption. The Dirichlet prior has been introduced to accommodate heterogeneity in the distributions of both errors (see \hyperlink{Chigira and Shiba, 2015}{Chigira and Shiba, 2015} for an example) and model parameters (\hyperlink{Allenby et al., 1998}{Allenby et al., 1998}) by mixing a fixed number of normal distributions.

A notable drawback of the Dirichlet prior is that the the dimension of the mixing distribution is usually unknown. Bayesian semi-parametric methods introduce more flexibility by letting the data and the prior determine the structure of heterogeneity jointly. The Dirichlet Process ($\mathcal{DP}$) prior\footnote{See \hyperlink{Escobar and West, 1995}{Escobar and West, 1995} and \hyperlink{Escobar and West, 1998}{1998} and \hyperlink{MacEachern, 1998}{MacEachern, 1998} for a reference of the Dirichlet Process prior.} can be used to form a mixing of normal distributions, whose dimension need not be predetermined. In this sense, the use of $\mathcal{DP}$ priors represents a more flexible approach to accommodating heterogeneity than the mixing of a fixed number of normal distributions with the Dirichlet prior.

In the context where heterogeneity in the distributions of errors is a major concern, $\mathcal{DP}$ priors are introduced for the  distributional parameters of the errors in the model\footnote{E.g. for each error vector to have its distinct normal distribution, its distributional parameters are a mean vector and a covariance matrix.}. Under this prior, the distributional parameters are put into groups, and assigned a group specific value. As a result, the corresponding errors will have group specific distributions.

A landmark study in this area is \hyperlink{Conley et al., 2008}{Conley et al. (2008)}, 
who introduce a Bayesian semi-parametric approach to the instrumental variable problem in a two stage least square type model. Due to the endogeneity of some explanatory variables, the errors in the two stages are correlated by construction. Instead of assuming that the joint errors in the two stages have an identical bivariate normal distribution (c.f., \hyperlink{Chao and Phillips, 1998}{Chao and Phillips, 1998}; \hyperlink{Geweke, 1996}{Geweke, 1996}; \hyperlink{Kleibergen and van Dijk, 1998}{Kleibergen and van Dijk, 1998}; \hyperlink{Rossi et al., 2005}{Rossi et al., 2005}), the authors introduce a Dirichlet process prior for the distributional parameters. This provides a semi-parametric version of two stage least squares, where the errors of the two stages jointly follow a non-parametric mixture of normal distributions.

In this paper we focus on relaxing the identical distribution assumption on the errors, but in a different scenario from that of \hyperlink{Conley et al., 2008}{Conley et al. (2008)}. We propose a semi-parametric Bayesian \textsc{gls} that incorporates the $\mathcal{DP}$ prior. The motivation is to incorporate more information in the error distribution by allowing their distributional parameters to differ across observations. The resulting distribution of the error terms will involve a mixture of normal distributions where the number of the normal components is influenced by both the prior and the data. We then introduce two specific cases of semi-parametric Bayesian \textsc{gls}, namely for equation systems and panel data.

The rest of the paper is organized as follow. In Section \ref{sec:repH} we briefly review the literature on the Dirichlet process and its application as a prior for semi-parametric Bayesian estimators. We then consider two special cases of the \textsc{gls}. The \textsc{sur} with the $\mathcal{DP}$ prior (\textsc{dp-sur}) is introduced in Section \ref{sec:SemiSUR} with simulation and empirical results. Section \ref{sec:DP-REM} motivates and introduces our semi-parametric Bayesian \textsc{gls} estimator for the random effects model with the $\mathcal{DP}$ prior (\textsc{dp-rem}) for panel data, together with simulation and empirical results. Section \ref{sec:Conclusion} concludes the paper.

\section{Bayesian GLS with Dirichlet Process Prior} \label{sec:repH}
In this section we introduce the generic form of Bayesian \textsc{gls} with the $\mathcal{DP}$ prior. In Section \ref{sec: BGLS review} we briefly review the literature in related areas. In Section \ref{sec: DP-GLS} we introduce the semi-parametric Bayesian \textsc{gls}.

\subsection{Literature Review} \label{sec: BGLS review}
Bayesian attempts to incorporate heterogeneity in the distributional parameters of the errors in the linear regression model can be traced back to \hyperlink{Geweke, 1993}{Geweke (1993)} with the use of an inverse gamma prior for the variances of the errors. He demonstrates that such a scale mixture of normal distributions is equivalent to the errors having a t-distribution.

Although the model with the t-distributed errors is more flexible than assuming a normal distribution for the errors, this approach depends upon the assumption that the normal distributions are mixed with inverse gamma distributed variances. As pointed out by \hyperlink{Koop, 2003}{Koop (2003)}, relaxing this assumption results in more flexible models, given that the errors are no longer restricted to having a t-distribution. This can be done by using a Dirichlet prior, the conjugate prior of a multinomial distribution, to mix a finite number of normal distributions.

The Dirichlet mixture model has emerged as a widely applied methodology for capturing heterogeneity in both linear and non-linear models, c.f., \hyperlink{Allenby et al., 1998}{Allenby et al. (1998)}, \hyperlink{Li and Tobias, 2011}{Li and Tobias (2011)} and \hyperlink{Chigira and Shiba, 2015}{Chigira and Shiba (2015)}. The main limitation is that it takes a fairly difficult test procedure to determine the ``correct'' number of mixing components.

In the wake of this limitation of the Dirichlet mixture model, it seems more reasonable to let the data and the prior jointly determine the number of normal components in the mixture. This can be achieved using a Dirichlet prior of infinite dimension, which is the Dirichlet process ($\mathcal{DP}$) introduced by \hyperlink{Ferguson, 1973}{Ferguson (1973)}\footnote{See \hyperlink{Teh, 2011}{Teh (2011)} and \hyperlink{Gershman and Blei, 2012}{Gershman and Blei (2012)} for reviews of the Dirichlet process.}. $\mathcal{DP}$ is the conjugate prior for a non-parametric multinomial distribution of infinite dimensions. The generic form of the $\mathcal{DP}$ can be written as
\begin{equation}
F \sim \mathcal{DP} \left(\alpha, F_{0} \right),
\end{equation}
where $\alpha>0$ is the concentration parameter, and $F_0$ is the base distribution. $F$ is a random distribution that is discrete with probability one\footnote{The level of discreteness is influenced by $\alpha$, the concentration parameter.}.

The $\mathcal{DP}$ is a non-parametric ``distribution of distributions'' (\hyperlink{Escobar and West, 1995}{Escobar and West, 1995} and \hyperlink{Escobar and West, 1998}{1998}; \hyperlink{MacEachern, 1998}{MacEachern, 1998}), in the sense that a draw, say $F$, from a $\mathcal{DP}$ is a probability distribution itself. Conditional on $n-1$ existing realizations $\{ r_1, r_2, \ldots, r_{n-1} \}$ from $F$, the Chinese restaurant process (\hyperlink{Aldous, 1985}{Aldous, 1985}) provides the predictive probabilities of the $n^{th}$ realization, $r_n$. Due to the fact that $F$ is discrete, the existing realizations will be assigned to groups, where all realizations in the same group take a group specific unique value.

Denote the group id of $r_i$ as $c_i = 1, \ldots, K$, and the unique value of group $c_i$ as $r_{c_i}^*$: if $r_i$ is in group $k$, then $r_i = r_{c_i}^* = r_{k}^*$. The prediction probabilities of $r_n$ is given by
\begin{equation} \label{CRP}
\Pr \left\{ r_n = r^*_k | r_1, r_2, \ldots, r_{n-1} \right\}
= \left\{
\begin{array}{ll}
\frac{n_k}{n-1+\alpha} & \text{if} \ 1 \leq k \leq K \\
& \\
\frac{\alpha}{n-1+\alpha} & \text{if} \ k = K+1 \ ( \text{i.e.,} \ r_n = r^*_{K+1} \sim F_0)
\end{array} \right.,
\end{equation}
where $n_k$ is the number of realizations that are already in group $k$. \hyperlink{Aldous, 1985}{Aldous (1985)} show that $r_1, r_2, \ldots, r_n$ generated according to the Chinese restaurant process are i.i.d. draws from $F$\footnote{The realisations $r_1, r_2, \ldots, r_n$ generated according to (\ref{CRP}) are not independent given that the $n^{th}$ realisation is generated conditioned on the $n-1$ realizations before. However, these realisations are exchangeable, and therefore independent conditional on a distribution $F$.}, i.e.,
\begin{equation} \label{dpx}
\begin{aligned}
F | \alpha, F_0 &\sim \mathcal{DP}(\alpha, F_{0}) \\
r_{i} | F &\overset{iid}{\sim} F.
\end{aligned}
\end{equation}

A model with a $\mathcal{DP}$ prior on the distribution of parameters is called a $\mathcal{DP}$ mixture model (c.f., \hyperlink{de Carvalho et al., 2013}{de Carvalho et al., 2013}; \hyperlink{Wiesenfarth et al. 2014}{Wiesenfarth et al., 2014}; \hyperlink{Li et al., 2018}{Li et al., 2018} and \hyperlink{Hejblum et al., 2019}{Hejblum et al., 2019}), and is capable of representing very general forms of heterogeneity in the distributions of the observations. The $\mathcal{DP}$ normal mixture model, whose mixture components are normal distributions, can be written as
\begin{equation} \label{dp normal mixture}
\begin{aligned}
F | \alpha, F_0 &\sim \mathcal{DP}(\alpha, F_{0}) \\
\theta_{i} | F &\overset{iid}{\sim} F \\
\bm{y}_i | \theta_i & \sim \mathcal{N} \left( \theta_i \right),
\end{aligned}
\end{equation}
where $\theta_i$ is the set of parameters of observation $\bm{y}_i$. In the multivariate normal\footnote{Note that $\bm{y}_i$ is a vector.} case, $\theta_i$ consists of the mean vector and covariance matrix, i.e., $\theta_i = \left( \bm{\mu}_i, \bm{\Sigma}_i \right)$.

The posterior probability of $\theta_i$ having the same value as one of the existing $\theta_{-i}$ is
\begin{equation} \label{ProbExistTheta}
\Pr \left\{\theta_{i} = \theta^*_{k} | \theta_{-i}, \bm{y}_i, \alpha \right\}
\propto
\frac{n_{k}}{n-1+\alpha}
f_{\mathcal{N}} \left( \bm{y}_i | \theta^*_{k} \right),
\end{equation}
where $\theta^*_{k}$ and $n_{k}$ respectively denote the unique value of group $k$  and the number of observations already in group $k$. $f_{\mathcal{N}} (\cdot)$ denotes the density function of multivariate normal distribution. The posterior probability of $\theta_i$ taking a new value $\theta^*_{new}$ from the base distribution is
\begin{equation} \label{ProbNewTheta}
\Pr \left\{\theta_{i} = \theta^*_{new} | \theta_{i}, \bm{y}_i, \alpha, F_0 \right\} \propto
\frac{\alpha}{n-1+\alpha}
\int f_{\mathcal{N}} \left( \bm{y}_i | \theta_{i} \right)
     f_{F_0} \left( \theta_{i} \right) \diff \theta_{i},
\end{equation}
where $f_{F_0} \left( \theta_{i} \right)$ is the probability density of the current value $\theta_{i}$ given $F_0$.

\subsection{Semi-parametric Bayesian GLS} \label{sec: DP-GLS}
In this section we introduce the generic form of the semi-parametric \textsc{gls} estimator, where a $\mathcal{DP}$ prior is introduced on the distributional parameters of the errors. Consider a general linear regression
\begin{equation} \label{Generic GLS}
\bm{y}_i = \bm{X}_i \bm{\beta} + \bm{\varepsilon}_i,
\end{equation}
where $i$ indexes the observation, $\bm{y}_i$ is a $Q \times 1$ vector of dependent variables, $\bm{X_i}$ is a $Q \times K$ matrix of explanatory variables, and $\bm{\beta}$ is a $K \times 1$ vector of coefficients. $\bm{\varepsilon}_i \sim \mathcal{N} \left( \theta_i \right)$ is a $Q \times 1$ vector of errors, where $\theta_i = \left( \bm{\mu}_i, \bm{\Sigma}_i \right)$ denotes the distribution parameters.

To facilitate a more flexible distribution for the errors, where both the mean and covariance matrix of each $\bm{\varepsilon_i}$ is allowed to be different, we do not include a constant in $\bm{X}_i$. Our semi-parametric \textsc{gls} estimator introduces a $\mathcal{DP}$ prior on the distribution of $\theta_i$, resulting in the errors having a non-parametric mixture of normal distributions. The hierarchical prior for $\theta_i $ is
\begin{equation} \label{dp prior GLS}
\begin{aligned}
F | \alpha, F_0 &\sim \mathcal{DP}(\alpha, F_{0}) \\
\theta_i | F &\overset{iid}{\sim} F,
\end{aligned}
\end{equation}
where $\alpha$ is the concentration parameter, and $F_0$ is the base distribution of the $\mathcal{DP}$, respectively.

Due to the discreteness of $F$ under the $\mathcal{DP}$ prior, the values of some $\theta_i$'s will be the same, thus putting them into the same group. The distribution parameters of $\bm{\varepsilon}_i$ can then be written as
\begin{equation}
\theta_i = \theta_{c_i}^* = \left( \bm{\mu}_{c_i}^*, \bm{\Sigma}_{c_i}^* \right)
\end{equation}
where $c_i$ denotes the group id of $\bm{\varepsilon}_i$, and the superscript $*$ denotes the group-specific values of parameters. If $c_i = c_j$, $i, j \in \{ 1, 2, \ldots, N \}$, $\bm{\varepsilon}_i$ and $\bm{\varepsilon}_j$ share the same group id and parameters, i.e., $\bm{\mu}_{c_i}^* = \bm{\mu}_{c_j}^*$ and $\bm{\Sigma}_{c_i}^* = \bm{\Sigma}_{c_j}^*$. Such ``grouping'' characteristic can help to reveal the structure of the unobserved heterogeneity in the data.

A straightforward choice of the base distribution is the normal-inverse Wishart distribution, i.e., $\theta \sim \mathcal{NIW} (\nu_0, \bm{W}_0, \bm{\lambda}_0, \kappa_0)$, or equivalently,
\begin{equation} \label{base distr}
\begin{aligned}
\bm{\Sigma} & \sim \mathcal{IW} (\nu_0, \bm{W}_0), \\
\bm{\mu} | \bm{\Sigma} & \sim \mathcal{N} \left( \bm{\lambda}_0,  \kappa_0^{-1} \bm{\Sigma} \right),
\end{aligned}
\end{equation}
where $\nu_0$, $\bm{W}_0$, $\bm{\lambda}_0$ and $\kappa_0$ are the hyper-parameters of the normal-inverse Wishart distribution. As the normal-inverse Wishart distribution is the conjugate prior for the parameters of a multivariate normal distribution, such a base distribution simplifies the evaluation of the posterior probability of $\theta_i$ taking a new value in (\ref{ProbNewTheta}). Having specified the $\mathcal{DP}$ prior, we now outline the process of  drawing from the posterior distributions of the parameters.

\subsection{MCMC Algorithm} \label{sec: DP GLS MCMC}
In order to take draws for the parameters in the model, the following Gibbs sampler is employed. We need to draw the distributional parameters $\Theta = \{ \theta_{c_i}^* \}_{i=1}^{N} = \{ \bm{\mu}_{c_i}^*, \bm{\Sigma}_{c_i}^* \}_{i=1}^{N}$, the regression parameters $\bm{\beta}$ and the concentration parameter $\alpha$ of the $\mathcal{DP}$ prior, i.e.,
\begin{equation} \label{MCMC GLS}
\begin{aligned}
\Theta & | \bm{y}, \bm{X}, \bm{\beta}, \alpha \\
\bm{\beta} & | \bm{y}, \bm{X}, \Theta, \alpha \\
\alpha & |  \bm{y}, \bm{X}, \bm{\beta}, \Theta.
\end{aligned}
\end{equation}

We begin with drawing the distributional parameters of the errors. For this we view the residuals $\bm{e}_i$ ($i = 1, \ldots, N$) as the ``observed'' errors.
According to (\ref{ProbExistTheta}) in the Chinese restaurant process, the probability of distributional parameters $\theta_i$ being assigned to group $c_j$ is

\begin{equation} \label{mcmc existing theta}
\tilde{p}_j =
\frac{1}{N - 1 + \alpha} f_{\mathcal{N}} \left( \bm{e}_i | \theta^*_{c_j} \right), \quad  \ j \neq i.
\end{equation}
Similarly by (\ref{ProbNewTheta}), the probability of $\theta_i$ being assigned to a new group, i.e., taking a new value drawn from $F_0$ is
\begin{equation} \label{mcmc new theta}
\tilde{p}_i =
\frac{\alpha}{N - 1 + \alpha}
\int f_{\mathcal{N}} \left( \bm{e}_i | \theta_i \right)
     f_{F_0}\left( \theta_i \right) \diff \theta_i.
\end{equation}

Normalizing $\tilde{p} = \{ \tilde{p}_1, \ldots, \tilde{p}_N \}$ gives us a multinomial probability vector
\begin{equation}
p = \frac{\tilde{p}}{\sum_{k = 1}^{N}{\tilde{p}_k}}, \quad \tilde{p} = \left[ \tilde{p}_1, \ldots, \tilde{p}_N \right]'.
\end{equation}
A draw can then be taken within $1, \ldots, N$ from the corresponding multinomial distribution to update the group membership. If the multinomial draw is $j \neq i$, $\theta_i$ is assigned to the group $c_j$, while if the draw is $i$, $\theta_i$ is assigned to a new group.

It should be mentioned that due to the choice of normal-inverse Wishart distribution for $F_0$, the integral in (\ref{mcmc new theta}) is
\begin{equation} \label{int NIW}
\int f_{\mathcal{N}} \left( \bm{e}_i | \theta_i \right)
     p\left( \theta_i| F_0 \right) \diff \theta_i =
     \frac{1}{\pi^{Q / 2}}
     \frac{\kappa_0^{Q/2} \Gamma_Q \left( \frac{\nu_0 + 1}{2} \right) |\bm{W_0}|^{\nu_0 / 2}}
     {\left( \kappa_0 + 1 \right)^{Q/2} \Gamma_Q \left( \frac{\nu_0}{2} \right) |\bm{W_0} + \bm{S}|^{\left(\nu_0 + 1\right) / 2}},
\end{equation}
where $Q$ is the dimension of the error term vector $\bm{\varepsilon}_i$, $\Gamma_Q(\cdot)$ is the multivariate gamma function, and
\begin{equation} \label{update W}
\bm{S} = \frac{\kappa_0}{\kappa_0 + 1} \left(\bm{e}_i - \bm{\lambda}_0 \right) \left(\bm{e}_i - \bm{\lambda}_0 \right)'.
\end{equation}
As for the value of the hyper-parameters, we follow \hyperlink{Conley et al., 2008}{Conley et al. (2008)} and set $\nu_0 = Q + 0.004$, $\bm{W}_0 = 0.17 \bm{I}$, $\bm{\lambda}_0 = \bm{0}$ and $\kappa_0 = 0.016$.

Once the group memberships of all the distributional parameters $\theta_i$ have been updated, the unique values of parameters for each group are redrawn (as suggested by \hyperlink{Escobar and West, 1998}{Escobar and West, 1998}). Let $N_k$ and $\bar{\bm{e}}_k$ be the number of residuals assigned to group $k$ and their sample mean, respectively. The unique values of this group $\theta_{k}^* = (\bm{\mu}_k^*, \bm{\Sigma}_k^*)$ is redrawn from a normal-inverse Wishart distribution denoted by $\mathcal{NIW} (\nu_k, \bm{W}_k, \bm{\lambda}_k, \kappa_k)$, whose parameters are

\begin{equation} \label{post NIW}
\begin{aligned}
\nu_k & = \nu_0 + N_k \\
\kappa_k & = \kappa_0 +N_k \\
\bm{\lambda}_k & = \kappa_k^{-1} \left(\kappa_0 \bm{\lambda}_0 + N_k \bar{\bm{e}} \right) \\
\bm{W}_k & = \bm{W}_0 + \bm{S}_k
				+ \frac{\kappa_0 N_k}{\kappa_0 + N_k} \left( \bar{\bm{e}}_k - \bm{\lambda}_0 \right) \left( \bar{\bm{e}}_k - \bm{\lambda}_0 \right)',
\end{aligned}
\end{equation}
where $\bm{S}_k = \sum_{c_i = k}{\left( \bm{e}_i - \bar{\bm{e}}_k \right) \left( \bm{e}_i - \bar{\bm{e}}_k \right)'}$.

After draws have been taken for the distributional parameters, we move on to the second part of the \textsc{mcmc} algorithm in \eqref{MCMC GLS}, which draws from the posterior of the regression parameters $\bm{\beta}$. The likelihood of $\bm{\beta}$ in (\ref{Generic GLS}) is
\begin{equation} \label{Generic GLS L}
p \left( \bm{y}_i | \bm{\beta}, \bm{\mu}_{c_i}^*, \bm{\Sigma}_{c_i}^* \right) =
\frac{1}{(2 \pi)^{Q/2}} |\bm{\Sigma}_{c_i}^*|^{-\frac{1}{2}}
\exp \left[ -\frac{1}{2} (\bm{y}_i - \bm{X}_i \bm{\beta} - \bm{\mu}_{c_i}^*)' \bm{\Sigma}_{c_i}^{*-1} (\bm{y}_i - \bm{X}_i \bm{\beta} - \bm{\mu}_{c_i}^*) \right].
\end{equation}
We specify a normal prior for $\bm{\beta}$, i.e.,
\begin{equation} \label{Generic GLS beta prior}
\bm{\beta} \sim \mathcal{N} (\bm{b}_0, \bm{V}_0),
\end{equation}
where $\bm{b}_0$ and $\bm{V}_0$ denote the prior mean and covariance matrix of $\bm{\beta}$, respectively. The posterior of $\bm{\beta}$ is
\begin{equation} \label{Generic GLS post}
\bm{\beta} | \bm{y},\bm{\mu}_{c_i}^*, \bm{\Sigma}_{c_i}^*  \sim \mathcal{N} (\bm{b}, \bm{V}),
\end{equation}
where
\begin{equation} \label{Generic GLS post var}
\bm{V} = \left( \bm{V}_0^{-1} + \sum_{i=1}^{N}{\bm{X}_i' \bm{\Sigma}_{c_i}^{*-1} \bm{X}_i} \right)^{-1},
\end{equation}
and
\begin{equation} \label{Generic GLS post mean}
\bm{b} = \bm{V} \left( \bm{V}_0^{-1} \bm{b}_0 + \sum_{i=1}^N {\bm{X}_i' \bm{\Sigma}_{c_i}^{*-1} (\bm{y}_i - \bm{\mu}_{c_i}^*)} \right).
\end{equation}
For the hyper-parameters we specify $\bm{b}_0 = \bm{0}$ and $\bm{V}_0 = 1000 \bm{I}_K$, in order to prevent a prior that is overly informative.

Finally, to make draws of the concentration parameter $\alpha$, we adopt the prior introduced by \hyperlink{Conley et al., 2008}{Conley et al. (2008)}, which is
\begin{equation}
p(\alpha) \propto \left(1 - \frac{\alpha - \alpha_{min}}{\alpha_{max} - \alpha_{min}} \right)^{\tau},
\end{equation}
where $\alpha_{min}$ and $\alpha_{max}$ are the pre-set lower and upper bounds of $\alpha$. Larger $\alpha$ leads to more groups being generated on average, i.e., the $\mathcal{DP}$ being less discrete. \hyperlink{Antoniak, 1974}{Antoniak (1974)} gives the distribution of $K$, the number of groups, conditioned on $\alpha$. The corresponding posterior of $\alpha$ is
\begin{equation}
p(\alpha | K) \propto \alpha^{K} \frac{\Gamma(\alpha)}{\Gamma(N+\alpha)} \left(1 - \frac{\alpha - \alpha_{min}}{\alpha_{max} - \alpha_{min}} \right)^{\tau}, \ \alpha \in \left(\alpha_{min}, \alpha_{max} \right).
\end{equation}
We set $\alpha_{min}$ to 0.1083 such that the mode of $K$ is 1, and set $\alpha_{max}$ to 1.834 so that the mode of $K$ is 5\% of the sample size\footnote{To test the sensitivity to the prior of $\alpha$, the hyper-parameter $\alpha_{max}$ has been adjusted so that the mode of $K$ is 10\% and 50\% of the sample size, respectively. In our experiments the results are not sensitive to these changes in $\alpha_{max}$.}. Following the suggestion of \hyperlink{Conley et al., 2008}{Conley et al. (2008)}, we set $\tau$ to 0.8.

\section{Semi-parametric Seemingly Unrelated Regression}\label{sec:SemiSUR}

In this section we introduce how the $\mathcal{DP}$ prior is incorporated with the \textsc{sur} for equation systems. Consider a system of $M$ equations
\begin{equation}\label{Two-eqn system}
\bm{y}_m = \bm{X}_m \bm{\beta}_m + \bm{\varepsilon}_m, \quad m = 1, \ldots, M,
\end{equation}
where $\bm{y}_m = \left[y_{mi}\right]_{m=1}^{M}$, $\bm{\varepsilon}_m = \left[\varepsilon_{mi}\right]_{m=1}^{M}$ are $N \times 1$ vectors of dependent variables and errors, respectively. $\bm{X}_m$ is an $N \times K_m$ matrix of explanatory variables. Following our assumptions in \ref{sec: DP-GLS}, there are no constants in the equations. $\bm{\beta}_m$ is a $K_m \times 1$ vector of parameters.

\subsection{DP Prior for SUR} \label{sec:DP prior for SUR}

In the presence of correlation between errors across the equations there exists an efficiency gain by utilising a system estimator. The \textsc{sur} (\hyperlink{Zellner, 1962}{Zellner, 1962}) was introduced for this task. The equations are stacked in the following way
\begin{equation} \label{Two-equation SUR model}
 \begin{bmatrix}
   \bm{y}_1 \\
   \vdots \\
   \bm{y}_M \\
 \end{bmatrix} =
\begin{bmatrix}
  \bm{X}_1 & \cdots & \bm{0} \\
  \vdots & \ddots & \vdots \\
  \bm{0} & \cdots & \bm{X}_M \\
\end{bmatrix}
\begin{bmatrix}
  \bm{\beta}_1 \\
  \vdots \\
  \bm{\beta}_M \\
\end{bmatrix} +
\begin{bmatrix}
  \bm{\varepsilon}_1 \\
  \vdots \\
  \bm{\varepsilon}_M \\
\end{bmatrix},
\end{equation}
or simply
\begin{equation}
\bm{y} = \bm{X \beta} + \bm{\varepsilon}.
\end{equation}

In comparison with the Bayesian \textsc{ols} estimator that usually assumes $\varepsilon_{mi} \overset{iid}{\sim} \mathcal{N}(\mu_m, \sigma_m^2)$ for all $m$, the errors $\bm{\varepsilon}_i$ are now identically multivariate normally distributed, i.e., $\bm{\varepsilon}_i \overset{iid}{\sim} \mathcal{N}(\bm{\mu}, \bm{\Sigma})$.
The covariance matrix of $\bm{\varepsilon}$ is then
\begin{equation} \label{Normal SUR cov}
\bm{\Omega} = \bm{\Sigma} \otimes \bm{I} =
\begin{bmatrix}
\sigma_{11} \bm{I_N} & \cdots & \sigma_{1M} \bm{I_N} \\
\vdots & \ddots & \vdots \\
\sigma_{M1} \bm{I_N} & \cdots & \sigma_{MM} \bm{I_N} \\
\end{bmatrix}, \ \text{s.t.} \ \sigma_{pq} = \sigma_{qp}, \ p, q = 1, \ldots, M,
\end{equation}
where ''$\otimes$" stands for the Kronecker product.

The \textsc{gls} type estimators (\textsc{sur} in this case) utilize the information in the covariance matrix to transform the data, so that the transformed errors are homoskedastic with no serial correlation. Such transformation is reflected in the likelihood as in (\ref{Generic GLS L}).\footnote{However, the covariance matrix $\bm{\Omega}$ has a specific form as in the \textsc{sur} in (\ref{Normal SUR cov}), instead of the general, positive definite symmetric form of a covariance matrix.} The prior and posterior of the parameters can be defined similarly to equations (\ref{Generic GLS beta prior}) to (\ref{Generic GLS post mean}).

Although \textsc{sur} accounts for the cross-equation correlation of errors, as \hyperlink{Wooldridge, 2003}{Wooldridge (2003)} has noted, the errors are assumed to be  identically distributed.  Moreover, unlike the frequentist \textsc{gls} estimator, this distribution is usually assumed to be normal in Bayesian methods. In this section we propose a new \textsc{dp-sur} estimator that makes no \textit{a priori} assumptions on the family of distribution of the errors. If we allow each observation $i$ to have its own distributional parameters, flexibility of the error distribution will lead to identification problems given cross sectional data. A compromise is to assign the observations into groups with the $\mathcal{DP}$ prior as in (\ref{dp prior GLS}), in which case the distribution parameters of $\bm{\varepsilon}_i$ of the \textsc{sur} in (\ref{Two-equation SUR model}) is
\begin{equation}\label{HetSurCov}
  \bm{\mu}_{c_i}^* =
  \begin{bmatrix}
  	\mu_{1, c_i}^* \\
  	\vdots \\
  	\mu_{M, c_i}^* \\  	
  \end{bmatrix}; \quad
  \bm{\Sigma}_{c_i}^* =
  \begin{bmatrix}
    \sigma_{11,c_i}^* & \cdots & \sigma_{1M,c_i}^* \\
    \vdots & \ddots & \vdots \\
    \sigma_{M1,c_i}^* & \cdots & \sigma_{MM,c_i}^* \\
  \end{bmatrix}, \
  s.t. \ \sigma_{pq,c_i}^* = \sigma_{qp,c_i}^*, \ p, q = 1, \ldots, M.
\end{equation}
The main difference from the parametric Bayesian \textsc{sur} is that the covariance matrix of each observation is now given in (\ref{HetSurCov}), which allows each group of observations to have its own unique values for the parameters. Then draws can be taken from the posteriors of parameters according to the Gibbs sampler described in Section \ref{sec: DP GLS MCMC}.

\subsection{A Simulation Experiment} \label{sec:SimExp}
In this section we conduct simulation experiments designed to evaluate and compare the performances of three estimators. The first one is our semi-parametric \textsc{dp-sur} in Section \ref{sec:DP prior for SUR}. The second is a Bayesian \textsc{sur}, where the errors have a parametric mixture of normal distributions with a Dirichlet prior (abbreviated as \textsc{dir-sur} hereafter) on the distributional parameters of the errors. The third is a parametric Bayesian \textsc{sur} where the errors have a normal distribution (abbreviated as \textsc{nor-sur} hereafter).

With not loss of generality, all simulation experiments are based on a two-equation system. Two explanatory variables are included in each equation, which are drawn from normal distributions with the following parameters
\begin{equation*} \label{X in simulation eq 1}
x_{11,i} \overset{iid}{\sim} \mathcal{N}(5, 2), \
x_{12,i} \overset{iid}{\sim} \mathcal{N}(6, 2),
\end{equation*}
and
\begin{equation*} \label{X in simulation eq 2}
x_{21,i} \overset{iid}{\sim} \mathcal{N}(-5, 2), \
x_{22,i} \overset{iid}{\sim} \mathcal{N}(6, 2).
\end{equation*}
We set the values of the parameters to $\bm{\beta}_1 = (1, -2)^{\prime}$ and  $\bm{\beta}_2 = (-1, 2)^{\prime}$.

We generate errors from two types of distributions to demonstrate our \textsc{dp-sur}. The first is the multivariate log-normal distribution, i.e., $\ln \bm{\varepsilon_i} \sim \mathcal{N} \left( \bm{0}, \bm{\Sigma} \right)$. The log-normal distribution is fat tailed with a positive mean, thus suitable for examining the performance of our semi-parametric \textsc{dp-sur}. For the covariance matrix of the bivariate normal distribution of $\ln \bm{\varepsilon}_i$, we specify its form as
\begin{equation*}
\bm{\Sigma} =
\begin{bmatrix}
\sigma^2 & 0.5 \sigma^2 \\
0.5 \sigma^2 & \sigma^2 \\
\end{bmatrix}.
\end{equation*}
Without loss of generality, we let the variances of $\varepsilon_{1i}$ and $\varepsilon_{2i}$ be identical, and fix the correlation between them at 0.5. In order to explore the performances of the estimators when extreme values are generated with different probabilities, we set $\sigma^2$ equal to three values: 1, 1.5 and 2.

The second type of error distribution is designed to demonstrate the performance of our \textsc{dp-sur} when the errors have multi-modal distributions. In order to adjust the heaviness of the tails of the distributions as well, we employ mixtures of multivariate Student t distributions, which are scale mixtures of multivariate normal distributions (\hyperlink{Andrews and Mallows, 1974}{Andrews and Mallows, 1974}). To avoid unnecessary complexity, we utilise a  mixture of two multivariate t distributions with the same degrees of freedom ($df$) and scale matrix, but with different means. More specifically, the scale matrix shared by both multivariate t distributions is
\begin{equation*}
\bm{\Sigma} =
\begin{bmatrix}
1 & 0.5 \\
0.5 & 1 \\
\end{bmatrix}.
\end{equation*}

Four $df$ are specified to adjust the heaviness of the tails of the mixture distribution, namely 2\footnote{We avoid the df of 1 because the multivariate t distribution has no well defined expectation under this circumstance.}, 4, 6 and $\infty$, with the tails becoming less heavy. Note that when the $df$ is $\infty$, the mixture distribution reduces to a mixture of multivariate normal distributions. The mean vectors of the two multivariate t are $[-1, -1]'$ and $[4, 4]'$. They are chosen to ensure that the mean vectors of the two mixture components are located far enough from one another, so that the mixture distributions are bi-modal for all four $df$. The two multivariate t distributions are allocated mixture weights of 0.4 and 0.6, respectively, resulting in the mixture distributions being asymmetric.

Three sample sizes are chosen for the simulation experiments: 100, 200, and 300. For each sample size, 100 samples are generated. In the following section we will report the average posterior means, standard deviations and mean squared errors over the 100 samples.

\subsection{DP-SUR Simulation Results} \label{sec:SimRes}
In this section we present the results of the simulations for the three estimators \textsc{dp-sur}, \textsc{dir-sur} and \textsc{nor-sur}.

\subsubsection{Results with Log-normal Distributions}
In Table \ref{tab: sur sim results Log-normal} we report the posterior means, standard deviations and the mean squared errors of the parameters averaged over the 100 samples, estimated with the three estimators when the errors are multivariate log-normal. The columns are divided into three blocks for the three sample sizes, each containing the results of the three estimators. The rows are also divided into three blocks for the three values of the variance $\sigma^2$. Each block contains the posterior mean, standard deviation and mean squared errors of the parameters. It should be noted that the \textsc{dir-sur} uses a fixed number of mixture components, which in our simulation studies is set to the posterior mode of the number of clusters given by the \textsc{dp-sur} averaged over the 100 samples.
\begin{table}[htbp]
  \centering
  \scriptsize
  \setstretch{1.5}
  \caption{Simulation Results for \textsc{SUR} with Log-normal Distributions}
    \begin{tabular}{ccccccccccccc}
    \hline
    \hline
          & $N$   & \multicolumn{3}{c}{100} &       & \multicolumn{3}{c}{200} &       & \multicolumn{3}{c}{300} \bigstrut\\
    \hline
          & Estimator & DP    & DIR   & NOR   &       & DP    & DIR   & NOR   &       & DP    & DIR   & NOR \bigstrut\\
    \hline
    \hline
          & Parameter &       &       &       &       &       & $\sigma^2=1$ &       &       &       &       &  \bigstrut\\
    \hline
    \multirow{4}[2]{*}{Mean} & $\beta_{11}$ & 1.0036  & 1.0084  & 1.0050  &       & 1.0027  & 1.0032  & 1.0078  &       & 1.0004  & 1.0000  & 0.9972  \bigstrut[t]\\
          & $\beta_{12}$ & -1.9934  & -1.9947  & -1.9964  &       & -1.9943  & -1.9963  & -1.9927  &       & -1.9987  & -1.9978  & -1.9957  \\
          & $\beta_{21}$ & -1.0039  & -1.0002  & -0.9972  &       & -1.0037  & -1.0089  & -1.0152  &       & -1.0022  & -1.0085  & -1.0127  \\
          & $\beta_{22}$ & 2.0026  & 2.0056  & 2.0060  &       & 1.9999  & 1.9991  & 1.9962  &       & 2.0026  & 1.9995  & 1.9977  \bigstrut[b]\\
    \hline
    \multirow{4}[2]{*}{SD} & $\beta_{11}$ & 0.0254  & 0.0653  & 0.0865  &       & 0.0186  & 0.0603  & 0.0783  &       & 0.0128  & 0.0443  & 0.0561  \bigstrut[t]\\
          & $\beta_{12}$ & 0.0273  & 0.0676  & 0.0927  &       & 0.0173  & 0.0543  & 0.0707  &       & 0.0125  & 0.0423  & 0.0549  \\
          & $\beta_{21}$ & 0.0274  & 0.0754  & 0.1022  &       & 0.0162  & 0.0535  & 0.0692  &       & 0.0144  & 0.0484  & 0.0615  \\
          & $\beta_{22}$ & 0.0273  & 0.0733  & 0.1012  &       & 0.0179  & 0.0572  & 0.0771  &       & 0.0131  & 0.0435  & 0.0564  \bigstrut[b]\\
    \hline
    \multirow{4}[2]{*}{MSE} & $\beta_{11}$ & 0.0014  & 0.0080  & 0.0263  &       & 0.0007  & 0.0060  & 0.0145  &       & 0.0004  & 0.0037  & 0.0079  \bigstrut[t]\\
          & $\beta_{12}$ & 0.0016  & 0.0095  & 0.0220  &       & 0.0008  & 0.0053  & 0.0099  &       & 0.0003  & 0.0030  & 0.0067  \\
          & $\beta_{21}$ & 0.0013  & 0.0097  & 0.0227  &       & 0.0005  & 0.0046  & 0.0099  &       & 0.0004  & 0.0044  & 0.0093  \\
          & $\beta_{22}$ & 0.0017  & 0.0086  & 0.0219  &       & 0.0007  & 0.0048  & 0.0120  &       & 0.0003  & 0.0031  & 0.0062  \bigstrut[b]\\
    \hline
    \hline
          &       &       &       &       &       &       & $\sigma^2=1.5$ &       &       &       &       &  \bigstrut\\
    \hline
    \multirow{4}[2]{*}{Mean} & $\beta_{11}$ & 1.0020  & 1.0045  & 0.9845  &       & 1.0008  & 0.9972  & 0.9943  &       & 1.0004  & 1.0012  & 1.0026  \bigstrut[t]\\
          & $\beta_{12}$ & -1.9952  & -1.9884  & -1.9806  &       & -1.9967  & -2.0005  & -2.0042  &       & -1.9994  & -1.9968  & -2.0067  \\
          & $\beta_{21}$ & -1.0034  & -0.9948  & -0.9912  &       & -0.9988  & -1.0030  & -0.9893  &       & -0.9991  & -0.9974  & -0.9901  \\
          & $\beta_{22}$ & 2.0016  & 2.0002  & 1.9912  &       & 1.9970  & 1.9983  & 1.9954  &       & 1.9991  & 2.0069  & 2.0191  \bigstrut[b]\\
    \hline
    \multirow{4}[2]{*}{SD} & $\beta_{11}$ & 0.0276  & 0.0986  & 0.1661  &       & 0.0154  & 0.0709  & 0.1172  &       & 0.0126  & 0.0647  & 0.1026  \bigstrut[t]\\
          & $\beta_{12}$ & 0.0292  & 0.0999  & 0.1758  &       & 0.0169  & 0.0731  & 0.1283  &       & 0.0123  & 0.0614  & 0.1015  \\
          & $\beta_{21}$ & 0.0280  & 0.1051  & 0.1683  &       & 0.0159  & 0.0711  & 0.1177  &       & 0.0117  & 0.0598  & 0.0913  \\
          & $\beta_{22}$ & 0.0252  & 0.0949  & 0.1543  &       & 0.0172  & 0.0727  & 0.1227  &       & 0.0116  & 0.0574  & 0.0904  \bigstrut[b]\\
    \hline
    \multirow{4}[2]{*}{MSE} & $\beta_{11}$ & 0.0017  & 0.0144  & 0.0676  &       & 0.0005  & 0.0078  & 0.0296  &       & 0.0003  & 0.0058  & 0.0234  \bigstrut[t]\\
          & $\beta_{12}$ & 0.0018  & 0.0170  & 0.0825  &       & 0.0006  & 0.0082  & 0.0377  &       & 0.0003  & 0.0058  & 0.0197  \\
          & $\beta_{21}$ & 0.0017  & 0.0172  & 0.0613  &       & 0.0005  & 0.0082  & 0.0354  &       & 0.0003  & 0.0059  & 0.0188  \\
          & $\beta_{22}$ & 0.0014  & 0.0145  & 0.0618  &       & 0.0006  & 0.0084  & 0.0391  &       & 0.0003  & 0.0051  & 0.0214  \bigstrut[b]\\
    \hline
    \hline
          &       &       &       &       &       &       & $\sigma^2=2$ &       &       &       &       &  \bigstrut\\
    \hline
    \multirow{4}[2]{*}{Mean} & $\beta_{11}$ & 1.0021  & 1.0039  & 0.9854  &       & 1.0031  & 1.0051  & 1.0084  &       & 1.0010  & 1.0098  & 1.0142  \bigstrut[t]\\
          & $\beta_{12}$ & -1.9950  & -1.9876  & -1.9762  &       & -2.0010  & -1.9938  & -2.0108  &       & -1.9980  & -2.0010  & -2.0179  \\
          & $\beta_{21}$ & -1.0041  & -1.0165  & -1.0213  &       & -0.9991  & -0.9995  & -0.9855  &       & -1.0001  & -1.0075  & -1.0154  \\
          & $\beta_{22}$ & 2.0082  & 2.0205  & 2.0655  &       & 2.0028  & 2.0117  & 1.9899  &       & 2.0003  & 1.9995  & 1.9927  \bigstrut[b]\\
    \hline
    \multirow{4}[2]{*}{SD} & $\beta_{11}$ & 0.0263  & 0.1197  & 0.2767  &       & 0.0166  & 0.0989  & 0.2247  &       & 0.0117  & 0.0777  & 0.1600  \bigstrut[t]\\
          & $\beta_{12}$ & 0.0294  & 0.1249  & 0.3173  &       & 0.0172  & 0.0975  & 0.2358  &       & 0.0121  & 0.0755  & 0.1637  \\
          & $\beta_{21}$ & 0.0263  & 0.1265  & 0.2716  &       & 0.0168  & 0.0983  & 0.2135  &       & 0.0111  & 0.0767  & 0.1598  \\
          & $\beta_{22}$ & 0.0278  & 0.1276  & 0.2941  &       & 0.0151  & 0.0881  & 0.1959  &       & 0.0112  & 0.0749  & 0.1647  \bigstrut[b]\\
    \hline
    \multirow{4}[2]{*}{MSE} & $\beta_{11}$ & 0.0017  & 0.0207  & 0.3210  &       & 0.0006  & 0.0131  & 0.1264  &       & 0.0003  & 0.0092  & 0.0609  \bigstrut[t]\\
          & $\beta_{12}$ & 0.0021  & 0.0239  & 0.2620  &       & 0.0007  & 0.0135  & 0.1239  &       & 0.0003  & 0.0083  & 0.0683  \\
          & $\beta_{21}$ & 0.0014  & 0.0224  & 0.1950  &       & 0.0006  & 0.0136  & 0.1332  &       & 0.0003  & 0.0081  & 0.0690  \\
          & $\beta_{22}$ & 0.0017  & 0.0262  & 0.2690  &       & 0.0005  & 0.0117  & 0.0933  &       & 0.0002  & 0.0088  & 0.0834  \bigstrut[b]\\
    \hline
    \end{tabular}%
  \label{tab: sur sim results Log-normal}%
\end{table}

One may see from Table \ref{tab: sur sim results Log-normal} that all three estimators give posterior means that are very close to the true values of the parameters. This indicates that all three estimators provide good point estimators for the parameters.

Regarding the posterior standard deviations, it is not surprising that all three estimators provide smaller posterior deviations when the sample size increases for each of the three $\sigma^2$. Table \ref{tab: sur sim results Log-normal} also shows that the \textsc{dp-sur} posterior standard deviations are uniformly smaller than the \textsc{dir-sur} ones, which are in turn smaller than the \textsc{nor-sur} posterior standard deviations. It is worth noticing that when $\sigma^2$ gets larger, the ratio of the \textsc{dp-sur} posterior standard deviations to the \textsc{dir-sur} become smaller for all sample sizes. This indicates that through a non-parametric mixture of normal distributions, our \textsc{dp-sur} achieves increasingly less dispersed posteriors than the \textsc{dir-sur} that employs a parametric mixture when the ``true'' distribution of the errors is more skewed and heavy tailed.

The same phenomena are also observed comparing the \textsc{dp-sur} posterior standard deviations to those of the \textsc{nor-sur}. In addition, the posterior standard deviations of the \textsc{dp-sur} are rather similar across the three $\sigma^2$ for the same sample size, which demonstrates its capability to fit more skewed and heavy tailed error distributions without having to make the posterior more dispersed. In contrast, both parametric estimators, i.e., the \textsc{dir-sur} and the \textsc{nor-sur}, record considerable larger posterior standard deviations when $\sigma^2$ increases.

The mean squared errors (\textsc{mse}) are one of the most important measures for the performance of estimators. Results in Table \ref{tab: sur sim results Log-normal} show that the \textsc{mse} of all three estimators decrease as sample size increases for all values of $\sigma^2$. We also see from Table \ref{tab: sur sim results Log-normal} that the \textsc{dp-sur} again dominates the two parametric estimators \textsc{dir-sur} and \textsc{nor-sur}, giving smaller \textsc{mse} under all circumstances. Similar to the behaviour of the posterior standard deviations\footnote{This is not surprising, for the mean squared error is the sum of the squared bias of the estimator and its variance. As all three estimators in our simulations achieved posterior means close to the true values of the parameters, the differences in their \textsc{mse} are mostly driven by their posterior standard deviations.}, for all sample sizes the ratios of the \textsc{dp-sur} \textsc{mse} to the \textsc{dir-sur} and \textsc{nor-sur} ones become smaller as $\sigma^2$ increases. Moreover, the \textsc{mse} of the \textsc{dp-sur} remain similar when $\sigma^2$ becomes larger, while those of the parametric \textsc{dir-sur} and \textsc{nor-sur} both increase considerably. This again indicates that our semi-parametric \textsc{dp-sur} has superior performance when the ``true'' distribution of the errors is fat tailed.

\subsubsection{Results with Mixed Multivariate t Distributions}
The results with the mixed multivariate t errors are presented in Table \ref{tab: SUR sim results mixed t}. The four horizontal blocks represent different $df$, i.e., 2, 4, 6 and $\infty$. They show that the posterior means given by all three estimators are close to the truths, indicating that they all give good point estimators for the parameters.

\begin{table}[htbp]
  \centering
  \scriptsize
  \setstretch{1.05}
  \caption{Simulation Results for \textsc{SUR} with Mixed t Distributions}
    \begin{tabular}{ccccccccccccc}
    \toprule
    \toprule
          & $N$   & \multicolumn{3}{c}{100} &       & \multicolumn{3}{c}{200} &       & \multicolumn{3}{c}{300} \\
    \midrule
          & Estimator & DP    & DIR   & NOR   &       & DP    & DIR   & NOR   &       & DP    & DIR   & NOR \\
    \midrule
    \midrule
          & Parameter &       &       &       &       &       & $df = 2$ &       &       &       &       &  \\
    \midrule
    \multirow{4}[2]{*}{Mean} & $\beta_{11}$ & 0.9877  & 0.9636  & 0.9454  &       & 0.9975  & 1.0080  & 0.9934  &       & 0.9957  & 0.9921  & 0.9911  \\
          & $\beta_{12}$ & -1.9780  & -1.9446  & -1.9190  &       & -2.0046  & -2.0006  & -2.0001  &       & -2.0016  & -1.9987  & -2.0069  \\
          & $\beta_{21}$ & -1.0099  & -1.0233  & -1.0375  &       & -1.0000  & -1.0060  & -1.0003  &       & -1.0071  & -1.0349  & -1.0363  \\
          & $\beta_{22}$ & 2.0014  & 2.0062  & 2.0145  &       & 2.0016  & 1.9974  & 1.9934  &       & 2.0097  & 2.0002  & 1.9855  \\
    \midrule
    \multirow{4}[2]{*}{S.D.} & $\beta_{11}$ & 0.0704  & 0.0988  & 0.1378  &       & 0.0483  & 0.0783  & 0.1098  &       & 0.0420  & 0.0695  & 0.0974  \\
          & $\beta_{12}$ & 0.0700  & 0.1002  & 0.1444  &       & 0.0481  & 0.0765  & 0.1105  &       & 0.0377  & 0.0617  & 0.0871  \\
          & $\beta_{21}$ & 0.0824  & 0.1209  & 0.1605  &       & 0.0478  & 0.0773  & 0.1004  &       & 0.0442  & 0.0719  & 0.0962  \\
          & $\beta_{22}$ & 0.0722  & 0.1060  & 0.1436  &       & 0.0532  & 0.0824  & 0.1103  &       & 0.0418  & 0.0667  & 0.0909  \\
    \midrule
    \multirow{4}[2]{*}{MSE} & $\beta_{11}$ & 0.0109  & 0.0182  & 0.0573  &       & 0.0046  & 0.0103  & 0.0332  &       & 0.0036  & 0.0082  & 0.0225  \\
          & $\beta_{12}$ & 0.0110  & 0.0192  & 0.0573  &       & 0.0054  & 0.0088  & 0.0290  &       & 0.0030  & 0.0058  & 0.0177  \\
          & $\beta_{21}$ & 0.0153  & 0.0253  & 0.0547  &       & 0.0045  & 0.0102  & 0.0221  &       & 0.0044  & 0.0094  & 0.0203  \\
          & $\beta_{22}$ & 0.0134  & 0.0220  & 0.0482  &       & 0.0070  & 0.0113  & 0.0233  &       & 0.0040  & 0.0078  & 0.0185  \\
    \midrule
    \midrule
          &       &       &       &       &       &       & $df = 4$ &       &       &       &       &  \\
    \midrule
    \multirow{4}[2]{*}{Mean} & $\beta_{11}$ & 1.0107  & 1.0286  & 1.0320  &       & 0.9992  & 0.9878  & 0.9844  &       & 0.9985  & 1.0023  & 1.0044  \\
          & $\beta_{12}$ & -1.9956  & -2.0183  & -2.0202  &       & -1.9955  & -2.0082  & -2.0133  &       & -2.0031  & -2.0018  & -2.0005  \\
          & $\beta_{21}$ & -1.0043  & -0.9873  & -0.9890  &       & -0.9961  & -0.9889  & -0.9887  &       & -0.9974  & -0.9853  & -0.9822  \\
          & $\beta_{22}$ & 2.0059  & 2.0327  & 2.0352  &       & 2.0031  & 2.0202  & 2.0246  &       & 2.0015  & 1.9997  & 1.9988  \\
    \midrule
    \multirow{4}[2]{*}{S.D.} & $\beta_{11}$ & 0.0699  & 0.0862  & 0.0887  &       & 0.0481  & 0.0597  & 0.0617  &       & 0.0366  & 0.0466  & 0.0481  \\
          & $\beta_{12}$ & 0.0650  & 0.0814  & 0.0843  &       & 0.0436  & 0.0556  & 0.0581  &       & 0.0349  & 0.0448  & 0.0467  \\
          & $\beta_{21}$ & 0.0683  & 0.0875  & 0.0902  &       & 0.0422  & 0.0539  & 0.0552  &       & 0.0364  & 0.0461  & 0.0474  \\
          & $\beta_{22}$ & 0.0682  & 0.0839  & 0.0872  &       & 0.0463  & 0.0588  & 0.0614  &       & 0.0378  & 0.0474  & 0.0493  \\
    \midrule
    \multirow{4}[2]{*}{MSE} & $\beta_{11}$ & 0.0119  & 0.0162  & 0.0182  &       & 0.0048  & 0.0069  & 0.0078  &       & 0.0032  & 0.0048  & 0.0056  \\
          & $\beta_{12}$ & 0.0102  & 0.0134  & 0.0146  &       & 0.0040  & 0.0055  & 0.0065  &       & 0.0025  & 0.0039  & 0.0043  \\
          & $\beta_{21}$ & 0.0095  & 0.0131  & 0.0150  &       & 0.0038  & 0.0057  & 0.0064  &       & 0.0027  & 0.0039  & 0.0044  \\
          & $\beta_{22}$ & 0.0096  & 0.0143  & 0.0166  &       & 0.0045  & 0.0062  & 0.0075  &       & 0.0031  & 0.0044  & 0.0049  \\
    \midrule
    \midrule
          &       &       &       &       &       &       & $df = 6$ &       &       &       &       &  \\
    \midrule
    \multirow{4}[2]{*}{Mean} & $\beta_{11}$ & 1.0028  & 0.9896  & 0.9889  &       & 1.0075  & 0.9945  & 0.9944  &       & 0.9993  & 1.0068  & 1.0065  \\
          & $\beta_{12}$ & -2.0018  & -2.0030  & -2.0050  &       & -2.0032  & -2.0003  & -2.0001  &       & -1.9998  & -1.9894  & -1.9900  \\
          & $\beta_{21}$ & -1.0089  & -1.0059  & -1.0085  &       & -0.9984  & -0.9878  & -0.9857  &       & -1.0006  & -1.0064  & -1.0064  \\
          & $\beta_{22}$ & 2.0051  & 2.0101  & 2.0116  &       & 1.9977  & 1.9861  & 1.9851  &       & 2.0060  & 2.0051  & 2.0059  \\
    \midrule
    \multirow{4}[2]{*}{S.D.} & $\beta_{11}$ & 0.0614  & 0.0748  & 0.0748  &       & 0.0422  & 0.0514  & 0.0517  &       & 0.0350  & 0.0423  & 0.0424  \\
          & $\beta_{12}$ & 0.0586  & 0.0731  & 0.0735  &       & 0.0404  & 0.0490  & 0.0495  &       & 0.0358  & 0.0430  & 0.0434  \\
          & $\beta_{21}$ & 0.0572  & 0.0681  & 0.0687  &       & 0.0433  & 0.0541  & 0.0542  &       & 0.0375  & 0.0450  & 0.0454  \\
          & $\beta_{22}$ & 0.0605  & 0.0719  & 0.0735  &       & 0.0411  & 0.0502  & 0.0505  &       & 0.0348  & 0.0419  & 0.0425  \\
    \midrule
    \multirow{4}[2]{*}{MSE} & $\beta_{11}$ & 0.0098  & 0.0128  & 0.0137  &       & 0.0037  & 0.0046  & 0.0047  &       & 0.0029  & 0.0034  & 0.0035  \\
          & $\beta_{12}$ & 0.0079  & 0.0103  & 0.0105  &       & 0.0033  & 0.0046  & 0.0048  &       & 0.0024  & 0.0033  & 0.0034  \\
          & $\beta_{21}$ & 0.0070  & 0.0091  & 0.0098  &       & 0.0036  & 0.0054  & 0.0057  &       & 0.0027  & 0.0040  & 0.0042  \\
          & $\beta_{22}$ & 0.0085  & 0.0106  & 0.0116  &       & 0.0036  & 0.0050  & 0.0052  &       & 0.0022  & 0.0030  & 0.0031  \\
    \midrule
    \midrule
          &       &       &       &       &       &       & $df = \infty$ &       &       &       &       &  \\
    \midrule
    \multirow{4}[2]{*}{Mean} & $\beta_{11}$ & 0.9995  & 1.0182  & 1.0187  &       & 1.0067  & 1.0198  & 1.0204  &       & 1.0055  & 1.0154  & 1.0154  \\
          & $\beta_{12}$ & -1.9989  & -2.0232  & -2.0222  &       & -2.0007  & -1.9914  & -1.9907  &       & -1.9953  & -1.9929  & -1.9927  \\
          & $\beta_{21}$ & -0.9895  & -0.9846  & -0.9842  &       & -1.0002  & -1.0018  & -1.0010  &       & -1.0034  & -1.0065  & -1.0064  \\
          & $\beta_{22}$ & 2.0047  & 2.0110  & 2.0103  &       & 2.0056  & 2.0052  & 2.0041  &       & 2.0008  & 2.0118  & 2.0115  \\
    \midrule
    \multirow{4}[2]{*}{S.D.} & $\beta_{11}$ & 0.0574  & 0.0683  & 0.0676  &       & 0.0388  & 0.0448  & 0.0449  &       & 0.0301  & 0.0352  & 0.0350  \\
          & $\beta_{12}$ & 0.0576  & 0.0660  & 0.0656  &       & 0.0384  & 0.0440  & 0.0443  &       & 0.0296  & 0.0356  & 0.0354  \\
          & $\beta_{21}$ & 0.0593  & 0.0705  & 0.0701  &       & 0.0359  & 0.0424  & 0.0421  &       & 0.0287  & 0.0337  & 0.0335  \\
          & $\beta_{22}$ & 0.0518  & 0.0602  & 0.0596  &       & 0.0393  & 0.0450  & 0.0450  &       & 0.0309  & 0.0361  & 0.0361  \\
    \midrule
    \multirow{4}[2]{*}{MSE} & $\beta_{11}$ & 0.0070  & 0.0093  & 0.0094  &       & 0.0035  & 0.0046  & 0.0047  &       & 0.0018  & 0.0026  & 0.0026  \\
          & $\beta_{12}$ & 0.0068  & 0.0085  & 0.0084  &       & 0.0031  & 0.0041  & 0.0042  &       & 0.0018  & 0.0027  & 0.0027  \\
          & $\beta_{21}$ & 0.0075  & 0.0097  & 0.0098  &       & 0.0028  & 0.0034  & 0.0034  &       & 0.0019  & 0.0025  & 0.0025  \\
          & $\beta_{22}$ & 0.0053  & 0.0067  & 0.0067  &       & 0.0034  & 0.0037  & 0.0038  &       & 0.0020  & 0.0027  & 0.0027  \\
    \bottomrule
    \end{tabular}%
  \label{tab: SUR sim results mixed t}%
\end{table}

The posterior standard deviations given by our semi-parametric \textsc{dp-sur} are smaller than those of the \textsc{dir-sur} and \textsc{nor-sur} under all circumstances. This demonstrates that our \textsc{dp-sur} produces posteriors that are considerably more concentrated than the two parametric estimators. It is worth noticing that for each sample size, the ratios of the \textsc{dp-sur} posterior standard deviations to those of the \textsc{dir-sur} and \textsc{nor-sur} are the smallest when $df = 2$, and increases with larger $df$. This is due to the tails of the multivariate t distributions mixed being heavier when $df$ is smaller. The same situation is observed with respect to the advantage of \textsc{dir-sur} over \textsc{nor-sur}. In addition, the \textsc{dir-sur} and \textsc{nor-sur} posterior standard deviations are almost the same when $df = \infty$, while \textsc{dp-sur} posterior standard deviations are still smaller, which indicates that the parametric mixture model experiences difficulties in identifying the heterogeneity in the error distribution.

Similar to the posterior standard deviations, the \textsc{mse} of the \textsc{dp-sur} are uniformly smaller than the \textsc{dir-sur} and \textsc{nor-sur} ones. This indicates that our semi-parametric \textsc{dp-sur} outperforms the parametric estimators when the error distribution is asymmetric and bi-modal. In addition, the advantages regarding \textsc{mse} of the \textsc{dp-sur} over the \textsc{dir-sur}/\textsc{nor-sur} are the largest when the $df$ is 2, and becomes smaller when the $df$ increases. That is, the semi-parametric \textsc{dp-sur} proves more efficient when the tails of the mixed multivariate t distribution are heavier. Moreover, one can observe that the \textsc{mse} of the \textsc{dir-sur} and the \textsc{nor-sur} are similar when $df = \infty$, while the \textsc{dp-sur} still has smaller \textsc{mse}. It demonstrates that the performance advantage of the \textsc{dir-sur} using a parametric mixture of normal distributions over the \textsc{nor-sur} diminishes faster than that of the \textsc{dp-sur} as the tails of the error distribution become less heavy.

\subsection{DP-SUR Empirical Examples} \label{sec:EmpEg}

In this section we apply our \textsc{dp-sur} estimator to the demand for factors of production with a generalized Leontief cost function (\hyperlink{Diewert, 1971}{Diewert, 1971}). The model is an equation system with as many equations as there are factors. We do not impose symmetry or homogeneity restrictions to make the model more general.

The dataset we use are from \hyperlink{Malikov et al. 2016}{Malikov et al. (2016)}, which contains 799 observations on 285 large U.S. banks in 2002, 2004 and 2006. The data includes quantities and prices of the inputs including labour, physical assets and borrowed funding, and the quantity of output, which is the loans made by a bank.

The equation system for the demands for factors is
\begin{equation} \label{Bank Funding}
\begin{aligned}
a_L & = \frac{L}{Y} = \beta_{LA} \frac{P_A}{P_L} + \beta_{LF} \frac{P_F}{P_L} + \beta_{LT} T + \varepsilon_{L}, \\
a_A & = \frac{A}{Y} = \beta_{AL} \frac{P_L}{P_A} + \beta_{AF} \frac{P_F}{P_A} + \beta_{AT} T + \varepsilon_{A}, \\
a_F & = \frac{F}{Y} = \beta_{FL} \frac{P_L}{P_F} + \beta_{FA} \frac{P_A}{P_F} + \beta_{FT} T + \varepsilon_{F},
\end{aligned}
\end{equation}
where $L$, $A$ and $F$ denote the quantity of labour, physical assets and borrowed funds, respectively; $T$ denotes the trend variable; $Y$ denotes output, and $P_k$ is the price of factor $k$, with $k \in \{L, A, F\}$. For  the errors we assume that conditional on the explanatory variables, $\left( \varepsilon_{Li}, \varepsilon_{Ai}, \varepsilon_{Fi} \right) \sim \mathcal{N} \left(\bm{\mu}_i, \bm{\Sigma}_i \right)$ where $\bm{\mu}_i$ and $\bm{\Sigma_i}$ are respectively the mean vector and covariance matrix of individual $i$. We allow the errors to be correlated across the three equations in the system for any particular individual, i.e., $\text{cov} \left( \varepsilon_{ki}, \varepsilon_{si} \right) \neq 0$, with $k, s \in \{L, A, F\}$ indexing equations.

With the generalized Leontief cost function, the cross price elasticities of the factors are given by
\begin{equation}
E_{ks} = \dfrac{1}{2} \dfrac{\beta_{ks} \left( P_k / P_s \right)^{-1/2}}{a_k}, \ \ \forall k \neq s.
\end{equation}
The own price elasticities are
\begin{equation}
E_{kk} = -\dfrac{1}{2} \dfrac{\sum_{s \neq k}{\beta_{ks} \left( P_k / P_s \right)^{-1/2}}}{a_k}.
\end{equation}

As the main interest in a demand system for factors of production is the price elasticities, we report results regarding the price elasticities of the three factors. There are nine elasticities within the three-input system, which are indexed by $\{L, A, F \}$ as stated before. In Table \ref{tab:Input Price Elasticities of U.S. Banks} we present the posterior means, standard deviations and p-values of the elasticities by the three estimators: \textsc{dp-sur}, \textsc{dir-sur} and \textsc{nor-sur}. In addition, we also demonstrate the performances of the three estimators with several figures.

\begin{table}[htbp]
  \centering
  \scriptsize
  \setstretch{1.5}
  \caption{Input Price Elasticities of U.S. Banks}
    \begin{tabular}{cccccccccccc}
    \toprule
    \toprule
          &       & Mean  &       &       &       & SD    &       &       &       & p-values     &  \\
    \midrule
    Parameter & DP    & Dir   & Nor   &       & DP    & Dir   & Nor   &       & DP    & Dir   & Nor \\
    \midrule
    $E_{LL}$ & -1.0286  & -0.8953  & -0.8932  &       & 0.0854  & 0.1208  & 0.1365  &       & 0.0000  & 0.0000  & 0.0000  \\
    $E_{LA}$ & 1.0331  & 0.8999  & 0.8977  &       & 0.0855  & 0.1210  & 0.1367  &       & 0.0000  & 0.0000  & 0.0000  \\
    $E_{LF}$ & -0.0045  & -0.0045  & -0.0045  &       & 0.0008  & 0.0011  & 0.0012  &       & 0.0000  & 0.0000  & 0.0000  \\
    $E_{AL}$ & 0.3072  & 0.3850  & 0.3834  &       & 0.0234  & 0.0386  & 0.0476  &       & 0.0000  & 0.0000  & 0.0000  \\
    $E_{AA}$ & -0.3011  & -0.3783  & -0.3769  &       & 0.0235  & 0.0388  & 0.0480  &       & 0.0000  & 0.0000  & 0.0000  \\
    $E_{AF}$ & -0.0061  & -0.0066  & -0.0066  &       & 0.0009  & 0.0015  & 0.0015  &       & 0.0000  & 0.0000  & 0.0000  \\
    $E_{FL}$ & 0.0540  & 0.0138  & 0.0118  &       & 0.0273  & 0.0302  & 0.0313  &       & 0.0420  & 0.6429  & 0.6857  \\
    $E_{FA}$ & -0.0434  & -0.0026  & -0.0013  &       & 0.0182  & 0.0196  & 0.0197  &       & 0.0160  & 0.8926  & 0.9526  \\
    $E_{FF}$ & -0.0107  & -0.0112  & -0.0105  &       & 0.0176  & 0.0208  & 0.0217  &       & 0.5400  & 0.5983  & 0.6097  \\
    \bottomrule
    \end{tabular}%
  \label{tab:Input Price Elasticities of U.S. Banks}%
\end{table}

From Table \ref{tab:Input Price Elasticities of U.S. Banks} one can see that the posterior means given by the three estimators are of the same signs and relatively close in magnitude for all nine elasticities. Among them, the posterior means of all three own-price elasticities are negative, with labour being the most elastic, followed by physical assets, and funding is the lease elastic input, which is expected with the banking industry. The cross-price elasticities demonstrate that labour and assets are substitutes, while assets and funding are complements. The relationship between labour and funding is not very clear: the demand for labour will decrease when the price for funding rises as indicated by $E_{LF}$, but $E_{FL}$ shows that the demand for funding will increase when labour becomes more expensive. However, it should be noted that both $E_{LF}$ and $E_{FL}$ are rather small in magnitude, showing the both are inelastic to changes in the price of the other.

The posterior standard deviations of our semi-parametric \textsc{dp-sur} are always smaller than the \textsc{dir-sur} or \textsc{nor-sur} ones. This difference is particularly significant with the cross-price elasticities regarding funding. Such a difference contributes to the fact that only the \textsc{dp-sur} posterior p-values are below the usually chosen significance level of 0.05 for $E_{FL}$ and $E_{FA}$.

\begin{figure}[h!]
 \centering
 \caption{95\% \textsc{hpdi}'s of Price Elasticities of the U.S. Banks' Factor Demands} \label{fig: CI-SUR}
 \includegraphics[width = .9\linewidth]{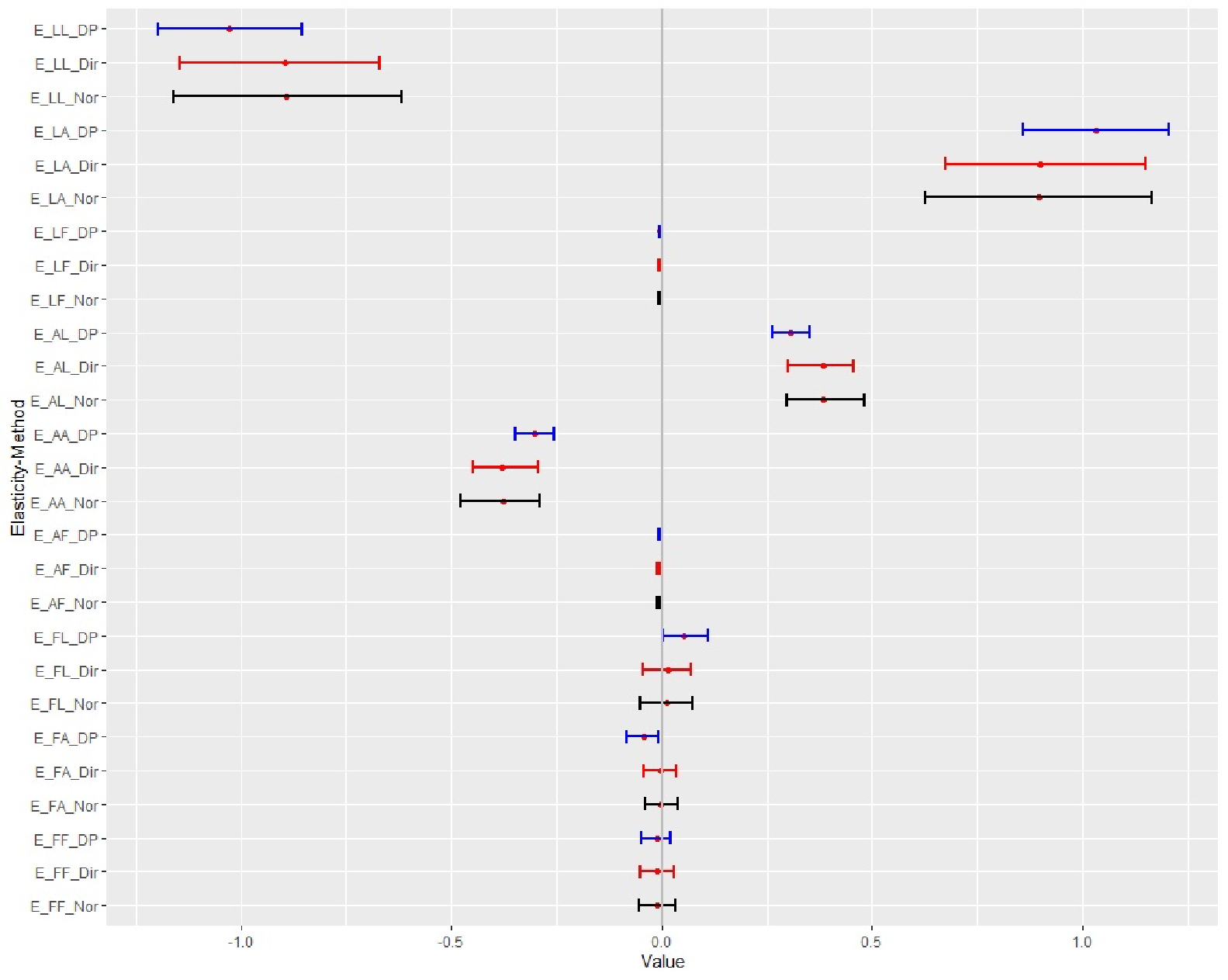}
 \begin{minipage}{0.8\linewidth}
    \footnotesize
    \setstretch{1.5}
    Note: We present all nine elasticities (e.g. ``E\_FA'' for the elasticity of funding w.r.t. the asset price) estimated with the three estimators: \textsc{dp-sur}, \textsc{dir-sur} and \textsc{nor-sur}. The three estimators are also allocated with the colours blue, red and black in the same order.
 \end{minipage}
\end{figure}

To better illustrate the performances of the three estimators, we present the 95\% highest posterior density intervals (\textsc{hpdi}'s) of the elasticities in the demand system of U.S. banks for factors in Figure \ref{fig: CI-SUR}. We observe that  the \textsc{hpdi}'s of \textsc{dp-sur} are the narrowest intervals for all nine elasticities. For instance, in the case of the elasticity of funding w.r.t. the asset price (\textsc{e\_fa}), only our \textsc{dp-sur} shows significance at the 5\% level.

\begin{figure}[h!]
\begin{center}
\caption{Histograms of the Elasticity of Funding w.r.t. Asset Price, U.S. Banks} \label{fig:Hist DP-SUR}
\begin{subfigure}{.3\textwidth}
  \centering
  \includegraphics[width=.8\linewidth]{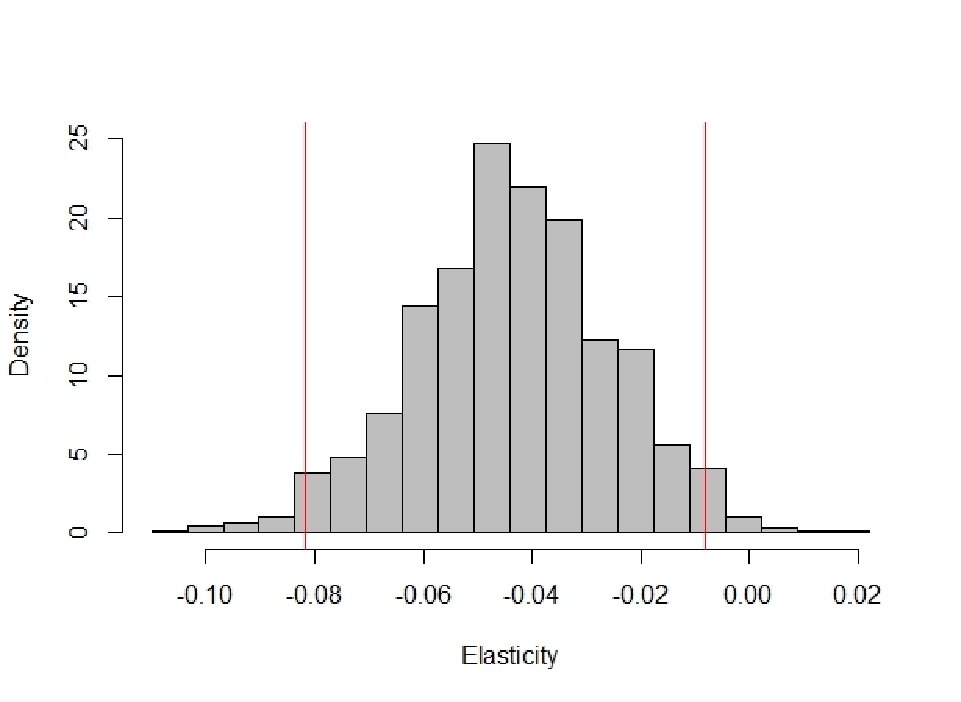}
  \caption{\textsc{dp-sur}}
  \label{fig: 1-a}
\end{subfigure}
\begin{subfigure}{.3\textwidth}
  \centering
  \includegraphics[width=.99\linewidth]{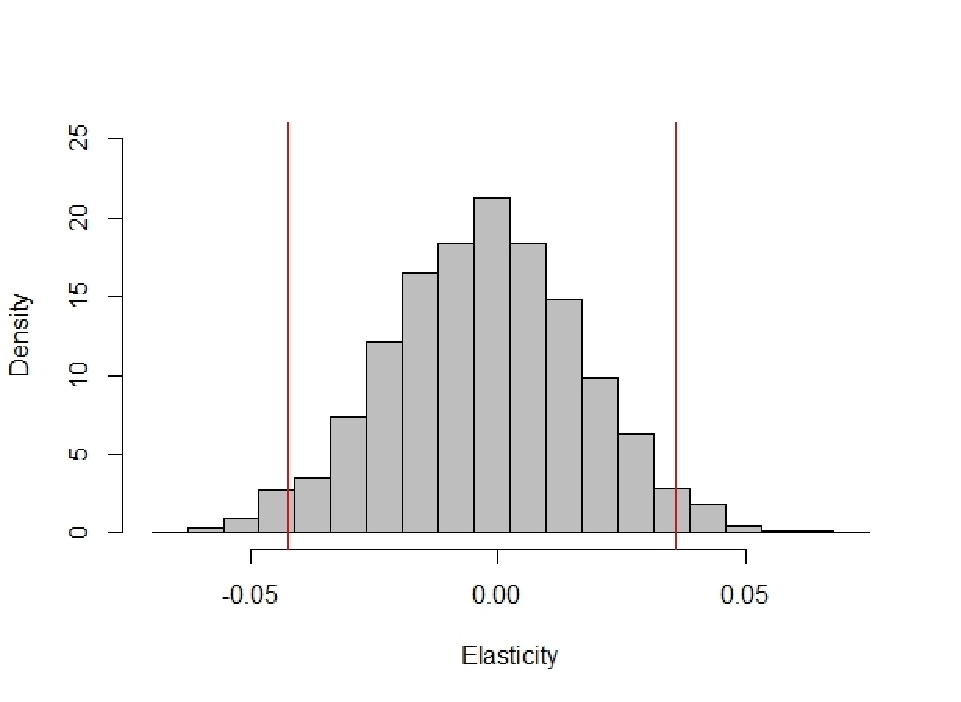}
  \caption{\textsc{dir-sur}}
  \label{fig: 1-b}
\end{subfigure}
\begin{subfigure}{.3\textwidth}
  \centering
  \includegraphics[width=.99\linewidth]{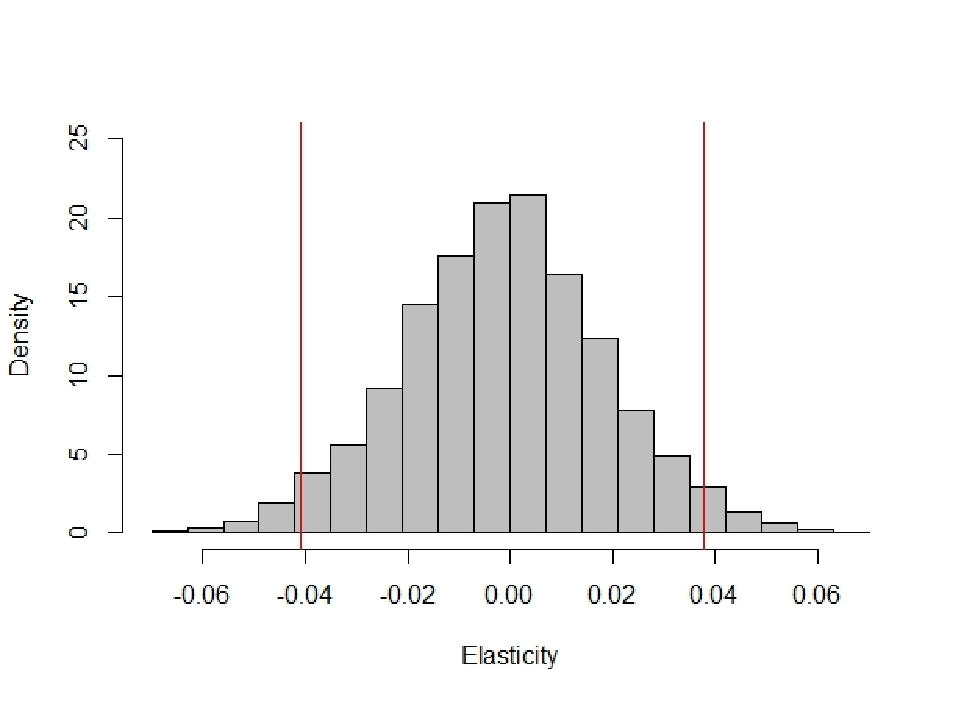}
  \caption{\textsc{nor-sur}}
  \label{fig: 1-c}
\end{subfigure}
\end{center}
\begin{center}
\begin{minipage}{0.8\linewidth}
  \footnotesize
  \setstretch{1.5}
Note: The three panels (\ref{fig: 1-a}), (\ref{fig: 1-b}) and (\ref{fig: 1-c}) are the histograms of the posterior draws for the elasticity of funding w.r.t. the price of assets by \textsc{dp-sur}, \textsc{dir-sur} and \textsc{nor-sur}, respectively. The two vertical lines (in red colour) mark the lower and upper bounds of the 95\% \textsc{hpdi}'s, which are $(-0.0818, -0.0082)$ for \textsc{dp-sur}, $(-0.0424, 0.0360)$ for \textsc{dir-sur}, and $(-0.0408, 0.0379)$ for \textsc{nor-sur}.
\end{minipage}
\end{center}
\end{figure}

We also present the histograms of $E_{FA}$ to further demonstrate the performance of the three estimators in Figure \ref{fig:Hist DP-SUR}. One may see that only the \textsc{dp-sur} gives such an \textsc{hpdi} that excludes 0. From the histogram one may also see that the posterior given by our \textsc{dp-sur} is left skewed with the non-parametric mixture. In fact, the posterior mode of the number of clusters is 3, showing that heterogeneity is detected in the error distributions.

\begin{figure}[h!]
\begin{center}
\caption{Predictive Densities of Residuals, U.S. Banks} \label{fig:Pred Den DP-SUR}
\begin{subfigure}{0.45\textwidth}
  \centering
  \includegraphics[width=.99\linewidth]{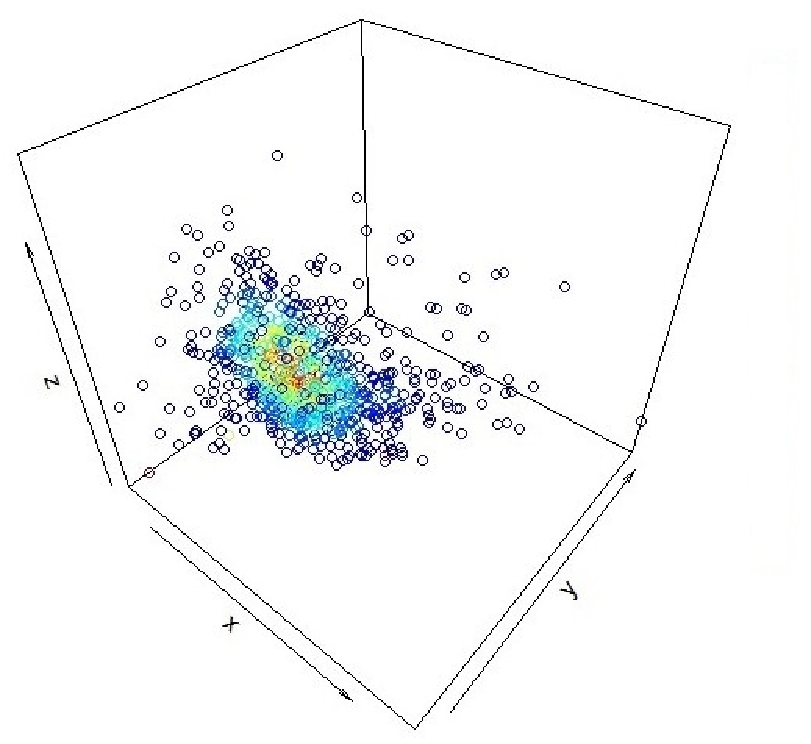}
  \caption{\textsc{dp-sur}}
  \label{fig: 2-a}
\end{subfigure}
\begin{subfigure}{0.45\textwidth}
  \centering
  \includegraphics[width=.99\linewidth]{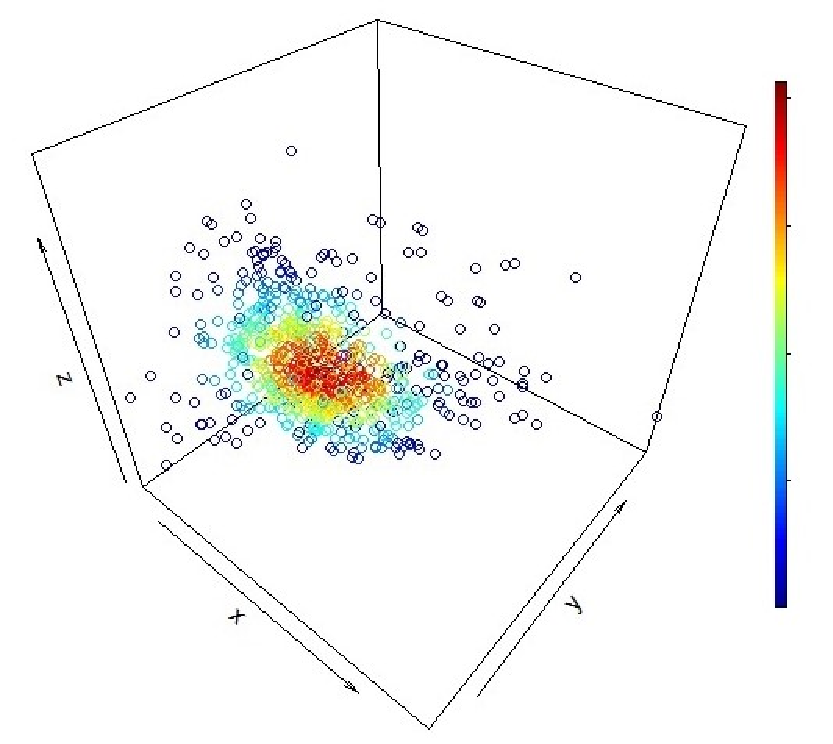}
  \caption{\textsc{nor-sur}}
  \label{fig: 2-b}
\end{subfigure}
\end{center}
\begin{center}
\begin{minipage}{0.8\linewidth}
\footnotesize
\setstretch{1.5}
Note: As there are three equations in the system, each residual is a 3 by 1 vector, which is represented by a circle in the 3D coordinate system. The x, y and z axes represent the residuals in the labour ($L$), assets ($A$) and funding ($F$) equations in \eqref{Bank Funding}, respectively. Predictive densities of the residuals are reflected in the colours of the circles. The circle colours change from blue to red with higher predictive densities of the corresponding residuals. Panels \ref{fig: 2-a} and \ref{fig: 2-b} show the residuals of \textsc{dp-sur} and \textsc{nor-sur}, respectively.
\end{minipage}
\end{center}
\end{figure}

In Figure \ref{fig:Pred Den DP-SUR} we present the posterior means of predictive densities of the residuals obtained by the \textsc{dp-sur} and \textsc{nor-sur}. One may see that with the semi-parametric \textsc{dp-sur}, the predictive density of the residuals is bi-modal, which is not captured by the \textsc{nor-sur}.

\section{Semi-parametric Approach to Random Effects Model} \label{sec:DP-REM}
The \textsc{gls} has also seen numerous applications with panel data models, and in particular the random effects model (\textsc{rem}). In a panel with $N$ units and $T$ time periods, the error of each unit\footnote{We use the term ``unit'' to denote the cross section here. In practice it can be households, firms, countries or persons.}
is a $T \times 1$\footnote{That is, the $Q$ in the generic semi-parametric \textsc{gls} in Section \ref{sec: DP-GLS} is $T$ in this context. We use $T$ here following panel data protocols.} vector. We will relax the assumption of parametric Bayesian \textsc{gls} for the \textsc{rem} (\hyperlink{Koop, 2003}{Koop, 2003}) that the error vectors for all units have the same distribution.  In this section we propose a semi-parametric Bayesian approach by introducing $\mathcal{DP}$ priors on the distributional parameters of the random effects and idiosyncratic errors. We follow the same approach as in the \textsc{dp-sur} method in terms of applying the $\mathcal{DP}$ prior on the distributional parameters.

Consider the following panel data model
\begin{equation} \label{REM}
y_{it} = \beta_1 x_{1it} + \cdots + \beta_K x_{Kit} + u_i + \eta_{it} = \bm{x}_{it}' \bm{\beta} + \varepsilon_{it},
\end{equation}
where $i$ and $t$ index the cross section and time series dimensions of the data, respectively. $y_{it}$ is the dependent variable, $\bm{x}_{it}$ denotes the explanatory variables, and  $\bm{\beta}$ is a conformable vector of parameters. $u_i$ is the time-invariant unobservable of unit $i$, and $\eta_{it}$ the idiosyncratic error term. $\varepsilon_{it} = u_i + \eta_{it}$ is the composite error.

In Bayesian methods the difference between the fixed and random effects lies in the choice of prior for the individual effects $u_i$: the fixed effects model assumes a non-hierarchical prior  for $u_i$; and the \textsc{rem} includes a hierarchical prior. The prior for $u_i$ in the \textsc{rem} can be written as
\begin{equation}
u_i | \mu_u, \sigma_u^2 \overset{iid}{\sim} \mathcal{N} \left(\mu_u, \sigma_u^2 \right),
\end{equation}
where $\mu_u$ and $\sigma_u^2$ are the mean and variance\footnote{Note that the distributional parameters of $u_i$ and $\eta_{it}$ are often assumed to be random, and have their own priors. However, for the moment we leave them fixed for the sake of simplicity.} of $u_i$, respectively. Assuming $\eta_{it} \overset{iid}{\sim} \mathcal{N} (\mu_{\eta}, \sigma_{\eta}^2)$, the likelihood of $\bm{\beta}$ marginalized over $u_i$ in the Bayesian \textsc{rem} may be written as

\begin{equation} \label{homo marginal likelihood}
p \left( \bm{y}_i | \bm{\beta}, \mu, \bm{\Sigma} \right) =
\frac{1}{(2 \pi)^{T/2}} |\bm{\Sigma}|^{-\frac{1}{2}}
\mathrm{exp} \left[ -\frac{1}{2} (\bm{y}_i - \bm{X}_i \bm{\beta} - \mu \iota_T)' \bm{\Sigma} (\bm{y}_i - \bm{X}_i \bm{\beta} - \mu \iota_T) \right],
\end{equation}
where $\bm{X}_i = \left[x_{1it}, \ldots, x_{Kit}\right]_{t=1}^{T}$ is a $T \times K$ matrix of explanatory variables, and $\bm{y}_i = [y_{it}]_{t=1}^{T}$ is a $T \times 1$ vector of dependent variables. $\mu = \mu_u + \mu_{\eta}$ is the mean of the composite error $\varepsilon_{it}$, and $\iota_T$ is a $T \times 1$ vector of ones. $\bm{\Sigma}$ is the covariance matrix of the $T \times 1$ composite error vector $\bm{\varepsilon}_i = [\varepsilon_{i1}, \varepsilon_{i2}, \ldots, \varepsilon_{iT} ]'$. Assuming the usual strict exogeneity in \textsc{rem}, $\bm{\Sigma}$ is

\begin{equation} \label{homo cov for epsilon}
\bm{\Sigma}
= \sigma_{\eta}^2 \bm{I}_{T} + \sigma_u^2 \iota_T \iota'_T =
\begin{bmatrix}
\sigma_{\eta}^2 + \sigma_u^2 & \sigma_u^2 & \cdots & \sigma_u^2 \\
\sigma_u^2 & \sigma_{\eta}^2 + \sigma_u^2 & \cdots & \sigma_u^2 \\
\vdots & \vdots & \ddots & \vdots \\
\sigma_u^2 & \sigma_u^2 & \cdots & \sigma_{\eta}^2 + \sigma_u^2  \\
\end{bmatrix}.
\end{equation}
%

\subsection{DP Prior for REM}
In this paper our primary focus is the  provision of more efficient inference by exploiting information in the heterogeneous distributions of unobservables. Thus, we maintain the approach of \textsc{gls}, and focus on heterogeneity in the distributional parameters of the unobservables instead of introducing heterogeneity to the distribution of model parameters themselves\footnote{
\hyperlink{Kleinman and Ibrahim, 1998}{Kleinman and Ibrahim (1998)} and \hyperlink{Kyung et al., 2010}{Kyung et al. (2010)} studied the heterogeneity in the model parameters across the units ($\bm{\beta_i}$). In contrast to our case, the $\mathcal{DP}$ prior was put on the parameters themselves in these papers.}.
In this sense, our method is in the same spirit as the literature pioneered by \hyperlink{Conley et al., 2008}{Conley et al. (2008)}. We relax the identical distribution assumptions for both $\eta_{it}$ and $u_i$ by introducing two independent $\mathcal{DP}$ priors on their distributional parameters\footnote{\hyperlink{Hirano, 2002}{Hirano (2002)} presents a semi-parametric autoregressive panel data model with individual effects. A $\mathcal{DP}$ prior is introduced on distributional parameters of idiosyncratic errors. Our estimator, although not considering dynamic panels, allows both the idiosyncratic errors and the individual effects to be heterogeneous in their distributions by introducing two independent $\mathcal{DP}$ priors for them, which leads to more flexibility.}.

The $\mathcal{DP}$ prior for $\theta_{\eta, it} = ( \mu_{\eta, it}, \sigma^2_{\eta, it})$, the parameters of the idiosyncratic error $\eta_{it}$, is
\begin{equation}
\begin{aligned}
G & \sim \mathcal{DP}\left(\alpha_\eta, G_{0} \right) \\
\theta_{\eta, it} | G & \sim G,
\end{aligned}
\end{equation}
where $\alpha_{\eta}$ and $G_0$ denote the concentration parameter and base distribution of the $\mathcal{DP}$ prior, respectively. As the prior is introduced across all idiosyncratic errors, the grouping of distributional parameters $\theta_{\eta, it}$ are not restricted to either within the cross section or within the time series dimensions. This means $\eta_{it}$ and $\eta_{is}$ can be allocated to different groups $c_{it}$ and $c_{is}$ such that they have different distributions with parameters $\theta_{\eta, c_{it}}^* = (\mu_{\eta, c_{it}}^*, \sigma_{\eta, c_{it}}^{2*})$ and $\theta_{\eta, c_{is}}^* = (\mu_{\eta, c_{is}}^*, \sigma_{\eta, c_{is}}^{2*})$, respectively. Similarly, $\eta_{it}$ and $\eta_{jt}$ can be in the same group, i.e., $c_{it} = c_{jt}$, making them identically distributed with parameters $\theta_{\eta, c_{it}}^* = \theta_{\eta, c_{jt}}^*$.

We write the $\mathcal{DP}$ prior for $\theta_{u, i} = (\mu_{u, i}, \sigma^2_{u, i})$, the parameters of the individual effects $u_i$'s,  as
\begin{equation} \label{DP prior on u}
\begin{aligned}
F & \sim \mathcal{DP}\left(\alpha_u, F_{0} \right) \\
\theta_{u, i} | F & \sim F,
\end{aligned}
\end{equation}
where $\alpha_u$ is the concentration parameter, and $F_0$ is the base distribution of the $\mathcal{DP}$ prior. Given that we maintain the assumption of strictly exogeneity of \textsc{rem}, it is reasonable to introduce a $\mathcal{DP}$ prior independent of $G$\footnote{For two mixtures of normal distributions, it is possible to introduce, e.g., a Hierarchical Dirichlet Process prior (\hyperlink{Teh et al., 2005}{Teh et al., 2005}) instead of two independent $\mathcal{DP}$ when dependence between the two priors is necessary. However, as we are considering the individual effects $u_i$ and idiosyncratic errors $\eta_{it}$ in the \textsc{rem} framework, it is plausible to introduce two independent $\mathcal{DP}$ priors in our case.} for the distributional parameters $\theta_{u, i}$.

The $\mathcal{DP}$ prior on the distributional parameters of individual effects $u_i$ generates groupings across the $N$ units. As a result, if $u_i$ and $u_j$ are in different groups $c_i$ and $c_j$, their parameters will take the values $\theta_{u, c_i}^* = (\mu_{u, c_i}^*, \sigma_{u, c_i}^{2*})$ and $\theta_{u, c_j}^* = (\mu_{u, c_j}^*, \sigma_{u, c_j}^{2*})$, respectively. This relaxes the \textsc{rem} assumption that the individual effects are identically distributed, as $u_i$ and $u_j$ are allowed to have different distributional parameters.

The mean vector of the composite error vector $\bm{\varepsilon}_i = u_i \iota + \bm{\eta}_i$ is then given by
\begin{equation}
\bm{\mu}_i = \mu_{u, c_i}^* \iota_T + \bm{\mu}_{\bm{\eta}_i}
= \mu_{u, c_i}^* \iota_T +
\begin{bmatrix}
\mu_{\eta, c_{i1}}^*, \ldots, \mu_{\eta, c_{iT}}^*
\end{bmatrix}',
\end{equation}
where $\bm{\mu}_{\bm{\eta}_i}$ is the mean vector of $\bm{\eta}_i$. The covariance matrix of $\bm{\varepsilon}_i$ is
\begin{equation} \label{hetero cov for composite error}
\bm{\Sigma}_i = \bm{\Sigma}_{\bm{\eta}_i} + \sigma_{u, c_i}^{2*} \iota_T \iota'_T
= \mathbf{diag} \left(\sigma_{\eta, c_{i1}}^{2*}, \ldots, \sigma_{\eta, c_{iT}}^{2*} \right) + \sigma_{u, c_i}^{2*} \iota_T \iota'_T,
\end{equation}
where $\bm{\Sigma}_{\bm{\eta}_i} = \mathbf{diag} \left(\sigma_{\eta, c_{i1}}^{2*}, \ldots, \sigma_{\eta, c_{iT}}^{2*} \right)$ is the diagonal covariance matrix of $\bm{\eta}_i$, where the diagonal elements are the variances of the idiosyncratic errors of unit $i$ over all time periods.

The \textsc{mcmc} algorithm utilised to draw from the posterior distributions is similar to that described in Section \ref{sec: DP GLS MCMC} with a few minor changes due to the particular form of the \textsc{dp-rem}. The most significant change is that now the individual effects $\bm{U} = \{u_i\}_{i=1}^{N}$ must also be drawn. In addition, there are two sets of distributional parameters, i.e., $\Theta_u$ and $\Theta_\eta$ for  $u_i$ and $\eta_{it}$, respectively. Similarly, there are concentration parameters for the two independent $\mathcal{DP}$ priors, namely $\alpha_u$ and $\alpha_{\eta}$. Accordingly, the Gibbs sampler consists of
\begin{equation}
\begin{aligned}
\Theta_\eta &| \bm{y}, \bm{X}, U, \bm{\beta}, \Theta_u, \alpha_u, \alpha_\eta \\
\Theta_u &| \bm{y}, \bm{X}, U, \bm{\beta}, \Theta_\eta, \alpha_u, \alpha_\eta \\
U &| \bm{y}, \bm{X}, \bm{\beta}, \Theta_u, \Theta_\eta, \alpha_u, \alpha_\eta \\
\bm{\beta} &| \bm{y}, \bm{X}, U, \Theta_u, \Theta_\eta, \alpha_u, \alpha_\eta \\
\alpha_u &| \bm{y}, \bm{X}, U, \bm{\beta}, \Theta_u, \Theta_\eta, \alpha_\eta \\
\alpha_\eta &| \bm{y}, \bm{X}, U, \bm{\beta}, \Theta_u, \Theta_\eta, \alpha_u.
\end{aligned}
\end{equation}
Among the parameters, the drawing of the concentration parameters $\alpha_u$ and $\alpha_{\eta}$ is exactly the same as described in Section \ref{sec: DP GLS MCMC}, because the two $\mathcal{DP}$ priors in our \textsc{dp-rem} are independent. The drawing of $\Theta_u$ and $\Theta_\eta$ are also similar to the procedure described in (\ref{mcmc existing theta}) to (\ref{post NIW}). It should be noted that the normal-inverse Wishart distribution in (\ref{base distr}) can still serve as the base distributions $G_0$ and $F_0$. However, as both $u_i$ and $\eta_{it}$ are scalars, they become the normal-inverse gamma distribution with both $\bm{\lambda}_0$ and $\bm{W}_0$ being scalars consequently.

The drawing of $\bm{U}$ is based on the fact that for a given $i$, the hierarchical prior for $u_i$ is
\begin{equation}
u_i | \mu_{u, c_i}^*, \sigma_{u, c_i}^{*2} \sim \mathcal{N} \left(\mu_{u, c_i}^*, \sigma_{u, c_i}^{*2} \right),
\end{equation}
which is the conjugate prior for the normal likelihood, leading to the posterior of $u_i$ being normal as well. The posterior variance of $u_i$ is
\begin{equation} \label{posterior variance, hetero u and eta}
s_{i}^2 = \left( \sigma_{u, c_i}^{*-2} + \iota'_T \bm{\Sigma}_{\bm{\eta}_i}^{-1} \iota_T \right)^{-1}.
\end{equation}
The posterior mean of $u_i$ is
\begin{equation} \label{posterior mean, hetero u and eta}
m_{i} = s_{i}^2 \left( \sigma_{u, c_i}^{*-2} \mu_{u, c_i}^* + \iota'_T \bm{\Sigma}_{\bm{\eta}_i}^{-1} \left(\bm{y}_i - \bm{X}_i \bm{\beta} - \bm{\mu}_{\bm{\eta}_i} \right) \right).
\end{equation}

As for the regression parameter $\bm{\beta}$, its likelihood marginalized over ${u_i}$ is given by
\begin{equation}
p \left( \bm{y}_i | \bm{\beta}, \bm{\mu}_i, \bm{\Sigma}_i \right) =
\frac{1}{(2 \pi)^{T/2}} |\bm{\Sigma}_i|^{-\frac{1}{2}}
\exp \left[ -\frac{1}{2} \left(\bm{y}_i - \bm{X}_i \bm{\beta} - \bm{\mu}_i \right)' \bm{\Sigma}_i^{-1} \left(\bm{y}_i - \bm{X}_i \bm{\beta} - \bm{\mu}_i \right) \right],
\end{equation}
which differs from (\ref{Generic GLS L}) only in that $\bm{\mu}_i$, the mean of the composite error $\bm{\varepsilon}_i$ is included. Compared with the marginal likelihood of the parametric Bayesian \textsc{rem} in (\ref{homo marginal likelihood}), the covariance matrix of the composite error vector $\bm{\varepsilon_i}$ is allowed to be different for each unit $i$ in the panel.

Given a conjugate normal prior for $\bm{\beta}$, i.e.,
\begin{equation*}
\bm{\beta} \sim \mathcal{N} (\bm{b}_0, \bm{V}_0),
\end{equation*}
where $\bm{b}_0$ and $\bm{V}_0$ respectively denote the prior mean and covariance matrix of $\bm{\beta}$, the posterior covariance matrix of $\bm{\beta}$, marginalized over $u_i$, is given by
\begin{equation}
\bm{V} = \left( \bm{V}_0^{-1} + \sum_{i=1}^{N}{\bm{X}_i' \bm{\Sigma}_i^{-1} \bm{X}_i} \right)^{-1}.
\end{equation}
The posterior mean vector is
\begin{equation}
\bm{b} = \bm{V} \left( \bm{V}_0^{-1} \bm{b}_0 + \sum_{i=1}^N {\bm{X}_i' \bm{\Sigma}_i^{-1} \left( \bm{y}_i - \bm{\mu}_i \right)} \right).
\end{equation}

A modified version of our \textsc{dp-rem} can be introduced in the spirit of the correlated \textsc{rem} introduced by \hyperlink{Mundlak, 1978}{Mundlak (1978)} and further discussed by \hyperlink{Chamberlain, 1982}{Chamberlain (1982)}, \hyperlink{Wooldridge, 2005}{Wooldridge (2005)}, \hyperlink{Murtazashvili and Wooldridge, 2008}{Murtazashvili and Wooldridge (2008)} and \hyperlink{Wooldridge, 2019}{Wooldridge (2019)}. This model offers a middle ground between the fixed and random effects models by allowing the individual effects to be correlated with $\bm{X}_i$ in a specific manner. This is achieved by specifying the individual effects as a linear function of the within unit means of the explanatory variables, i.e.,
\begin{equation} \label{CREM}
\begin{aligned}
y_{it} & =  \beta_1 x_{1it} + \cdots + \beta_K x_{Kit} + a_i + \eta_{it}, \\
a_i & = \gamma_1 \bar{x}_{1i} + \cdots + \gamma_K \bar{x}_{Ki} + u_i,
\end{aligned}
\end{equation}
where $\bar{x}_{ki} = 1/T \sum_{t=1}^{T}{x_{kit}}$. Then a $\mathcal{DP}$ prior as in (\ref{DP prior on u}) can be applied to $u_i$, and the analyses of the parameters are similar.

\subsection{DP-REM Simulation Results} \label{sec:DP-REM simulation}
We carry out a series of simulation experiments to demonstrate the performance of our \textsc{dp-rem}. Similar to the simulations for the \textsc{dp-sur}, our \textsc{dp-rem} is compared with two other estimators. The first one is a Bayesian \textsc{rem} with a Dirichlet prior on the distributional parameters of $u_i$ and $\eta_{it}$ (\textsc{dir-rem} hereafter), where $u_i$ and $\eta_{it}$ have parametric mixtures of normal distributions. The second one assumes that $u_i$ and $\eta_{it}$ are normal distributed (\textsc{nor-rem} hereafter).

The simulations are carried out using the following specification
\begin{equation} \label{REM sim reg}
y_{it} = \beta_1 x_{1it} + \beta_2 x_{2it} + u_i + \eta_{it},
\end{equation}
where the explanatory variables are generated from
\begin{equation*}
x_{1it} \sim \mathcal{N} \left(1, 1 \right), \quad x_{2it} \sim \mathcal{N} \left(3, 1 \right)
\end{equation*}
with
\begin{equation*}
\beta_1 = 5, \ \beta_2 = -5.
\end{equation*}

The individual effects $u_i$ and the idiosyncratic errors $\eta_{it}$ are independently generated.
 As with the simulation experiments for the \textsc{dp-sur}, we specify two types of distributions: a log-normal distribution and a bi-modal distributions. For the log-normal distribution, the variances of $\ln u_i$ and $\ln \eta_{it}$ are chosen to be equal and denoted by $\sigma^2$, which are set to three values: 1, 1.5 and 2, in order to explore the performance of our \textsc{dp-rem} under different circumstances.

To create a bi-modal distribution, we mix two non-central univariate t distributions with non-centrality parameters -1 and 4. Like in the experiment design in Section \ref{sec:SimExp}, the $df$ are set to 2, 4, 6 and $\infty$ to adjust the heaviness of the tails. The weights assigned to the two non-central t distributions are 0.4 and 0.6, so that the mixture distribution is asymmetric.

To explore the performance of our \textsc{dp-rem} estimator with different sample sizes, we fix the number of periods in the panel at 3 and set the number of cross sections to 100, 200 and 300. 100 samples are generated for each sample size, and the results reported in this section are the average over all the samples.

\subsubsection{Results with Log-normal Distributions}
Table \ref{tab: rem sim results lognormal} presents the results of the simulations with log-normal distributed individual effects $u_i$ and idiosyncratic errors $\eta_{it}$. For the posterior means we observe that the three estimators all perform well. When $\sigma^2 = 2$, we observe the greatest differences for the \textsc{dir-rem} and \textsc{nor-rem} posterior means from the true values. It shows that the two parametric methods are slightly inferior as point estimators when the distributions of $u_i$ and $\eta_{it}$ in (\ref{REM sim reg}) are more skewed and heavy tailed. This can be caused by the distribution of the composite errors being a convolution of two log-normal distributions, which are fat tailed. The two parametric estimators \textsc{dir-rem} and \textsc{nor-rem} struggle to give good point estimates for the parameters in this case, while our semi-parametric \textsc{dp-rem} uniformly performs well.

\begin{table}[htbp]
  \centering
  \scriptsize
  \setstretch{1.5}
  \caption{Simulation Results for \textsc{REM} with Log-normal Distributions}
    \begin{tabular}{ccccccccccccc}
    \hline
    \hline
          & $N$   & \multicolumn{3}{c}{100} &       & \multicolumn{3}{c}{200} &       & \multicolumn{3}{c}{300} \bigstrut\\
    \hline
          & Estimator & DP    & DIR   & NOR   &       & DP    & DIR   & NOR   &       & DP    & DIR   & NOR \bigstrut\\
    \hline
    \hline
          & Parameter &       &       &       &       &       & $\sigma^2=1$ &       &       &       &       &  \bigstrut\\
    \hline
    \multirow{2}[2]{*}{Mean} & $\beta_1$ & 5.0094  & 5.0043  & 5.0227  &       & 4.9905  & 4.9833  & 4.9905  &       & 4.9998  & 4.9918  & 4.9983  \bigstrut[t]\\
          & $\beta_2$ & -4.9933  & -5.0228  & -4.9999  &       & -4.9961  & -5.0232  & -5.0026  &       & -4.9992  & -5.0229  & -5.0054  \bigstrut[b]\\
    \hline
    \multirow{2}[2]{*}{SD} & $\beta_1$ & 0.0475  & 0.1279  & 0.1311  &       & 0.0313  & 0.0928  & 0.0941  &       & 0.0250  & 0.0786  & 0.0799  \bigstrut[t]\\
          & $\beta_2$ & 0.0424  & 0.1160  & 0.1229  &       & 0.0288  & 0.0846  & 0.0893  &       & 0.0235  & 0.0730  & 0.0758  \bigstrut[b]\\
    \hline
    \multirow{2}[2]{*}{MSE} & $\beta_1$ & 0.0047  & 0.0301  & 0.0362  &       & 0.0021  & 0.0157  & 0.0181  &       & 0.0014  & 0.0114  & 0.0131  \bigstrut[t]\\
          & $\beta_2$ & 0.0039  & 0.0248  & 0.0291  &       & 0.0021  & 0.0149  & 0.0172  &       & 0.0014  & 0.0102  & 0.0122  \bigstrut[b]\\
    \hline
    \hline
          &       &       &       &       &       &       & $\sigma^2=1.5$ &       &       &       &       &  \bigstrut\\
    \hline
    \multirow{2}[2]{*}{Mean} & $\beta_1$ & 5.0034  & 5.0012  & 5.0276  &       & 5.0015  & 4.9650  & 4.9666  &       & 4.9947  & 4.9868  & 4.9935  \bigstrut[t]\\
          & $\beta_2$ & -5.0010  & -5.0746  & -5.0115  &       & -4.9965  & -5.0356  & -4.9842  &       & -4.9948  & -5.0385  & -5.0001  \bigstrut[b]\\
    \hline
    \multirow{2}[2]{*}{SD} & $\beta_1$ & 0.0498  & 0.2120  & 0.2273  &       & 0.0348  & 0.1670  & 0.1806  &       & 0.0251  & 0.1266  & 0.1337  \bigstrut[t]\\
          & $\beta_2$ & 0.0473  & 0.1905  & 0.2178  &       & 0.0303  & 0.1408  & 0.1605  &       & 0.0230  & 0.1163  & 0.1282  \bigstrut[b]\\
    \hline
    \multirow{2}[2]{*}{MSE} & $\beta_1$ & 0.0055  & 0.0878  & 0.1254  &       & 0.0028  & 0.0471  & 0.0686  &       & 0.0013  & 0.0259  & 0.0345  \bigstrut[t]\\
          & $\beta_2$ & 0.0055  & 0.0758  & 0.1121  &       & 0.0022  & 0.0368  & 0.0596  &       & 0.0012  & 0.0273  & 0.0336  \bigstrut[b]\\
    \hline
    \hline
          &       &       &       &       &       &       & $\sigma^2=2$ &       &       &       &       &  \bigstrut\\
    \hline
    \multirow{2}[2]{*}{Mean} & $\beta_1$ & 4.9962  & 4.9449  & 5.0006  &       & 5.0103  & 4.9520  & 4.9802  &       & 4.9982  & 4.9382  & 4.9341  \bigstrut[t]\\
          & $\beta_2$ & -4.9927  & -5.1079  & -5.0329  &       & -4.9952  & -5.1090  & -5.0282  &       & -4.9996  & -5.0779  & -4.9864  \bigstrut[b]\\
    \hline
    \multirow{2}[2]{*}{SD} & $\beta_1$ & 0.0524  & 0.3312  & 0.3909  &       & 0.0342  & 0.2478  & 0.2999  &       & 0.0259  & 0.2039  & 0.2438  \bigstrut[t]\\
          & $\beta_2$ & 0.0467  & 0.2907  & 0.3735  &       & 0.0296  & 0.2091  & 0.2714  &       & 0.0232  & 0.1786  & 0.2284  \bigstrut[b]\\
    \hline
    \multirow{2}[2]{*}{MSE} & $\beta_1$ & 0.0058  & 0.1911  & 0.4124  &       & 0.0029  & 0.1035  & 0.2028  &       & 0.0012  & 0.0725  & 0.1296  \bigstrut[t]\\
          & $\beta_2$ & 0.0050  & 0.1561  & 0.2932  &       & 0.0019  & 0.0980  & 0.1878  &       & 0.0012  & 0.0635  & 0.1121  \bigstrut[b]\\
    \hline
    \end{tabular}%
  \label{tab: rem sim results lognormal}%
\end{table}

The posterior standard deviations of all three estimators decrease as the sample size increases for all three values of $\sigma^2$. Among them, the \textsc{dp-rem} posterior standard deviations are always smaller than \textsc{dir-rem} and \textsc{nor-rem} ones. This indicates that the semi-parametric \textsc{dp-rem} provides posterior distributions that are less dispersed than the two parametric methods, as a result of the non-parametric mixture of normal distributions. It can also be seen that the posterior standard deviations of our \textsc{dp-rem} remains similar when $\sigma^2$ increases for any given sample size. In comparison, the posterior standard deviations of the two parametric estimators increase considerably when $\sigma^2$ gets larger.\footnote{This result was also observed in the simulations for equation systems as in Table \ref{tab: sur sim results Log-normal}.}

The \textsc{mse} for the \textsc{dp-rem} are always smaller than their \textsc{dir-rem} and \textsc{nor-rem} counterparts. The superior performance of the \textsc{dp-rem} demonstrates the advantage of the non-parametric mixture with the $\mathcal{DP}$ prior over the parametric mixture using the Dirichlet prior, and the homogeneous normal errors. As with the posterior standard deviations, the \textsc{mse} of the \textsc{dp-rem} remain similar when $\sigma^2$ increases, while the \textsc{dir-rem} and \textsc{nor-rem} ones increase considerably. This further demonstrates the superiority of our semi-parametric \textsc{dp-rem} when the individual effects and idiosyncratic errors have fat tailed distributions.

\subsubsection{Results with Mixed t Distributions}
Table \ref{tab:tab: rem sim results t} reports the results with mixed t distributed individual effects and idiosyncratic errors. It can be seen that for all three estimators their posterior means are close to the truth in all settings. That is, all three estimators perform well as point estimators.


%
\begin{table}[htbp]
  \centering
  \scriptsize
  \setstretch{1.5}
  \caption{Simulation Results for \textsc{REM} with Mixed t Distributions}
    \begin{tabular}{ccccccccccccc}
    \hline
    \hline
          & $N$   & \multicolumn{3}{c}{100} &       & \multicolumn{3}{c}{200} &       & \multicolumn{3}{c}{300} \bigstrut\\
    \hline
          & Estimator & DP    & DIR   & NOR   &       & DP    & DIR   & NOR   &       & DP    & DIR   & NOR \bigstrut\\
    \hline
    \hline
          & Parameter &       &       &       &       &       & $df = 2$ &       &       &       &       &  \bigstrut\\
    \hline
    \multirow{2}[2]{*}{Mean} & $\beta_1$ & 4.9958  & 5.0267  & 5.0215  &       & 4.9894  & 5.0021  & 5.0209  &       & 4.9939  & 5.0195  & 5.0251  \bigstrut[t]\\
          & $\beta_2$ & -4.9824  & -4.9424  & -4.9473  &       & -5.0018  & -5.0058  & -5.0226  &       & -5.0022  & -5.0164  & -5.0128  \bigstrut[b]\\
    \hline
    \multirow{2}[2]{*}{S.D.} & $\beta_1$ & 0.0974  & 0.1777  & 0.1996  &       & 0.0679  & 0.1300  & 0.1583  &       & 0.0548  & 0.1036  & 0.1181  \bigstrut[t]\\
          & $\beta_2$ & 0.0869  & 0.1632  & 0.1881  &       & 0.0611  & 0.1190  & 0.1521  &       & 0.0544  & 0.1079  & 0.1263  \bigstrut[b]\\
    \hline
    \multirow{2}[2]{*}{MSE} & $\beta_1$ & 0.0216  & 0.0522  & 0.0940  &       & 0.0096  & 0.0270  & 0.0576  &       & 0.0067  & 0.0185  & 0.0337  \bigstrut[t]\\
          & $\beta_2$ & 0.0164  & 0.0494  & 0.0838  &       & 0.0081  & 0.0236  & 0.0584  &       & 0.0068  & 0.0190  & 0.0360  \bigstrut[b]\\
    \hline
    \hline
          &       &       &       &       &       &       & $df = 4$ &       &       &       &       &  \bigstrut\\
    \hline
    \multirow{2}[2]{*}{Mean} & $\beta_1$ & 4.9982  & 4.9865  & 4.9869  &       & 4.9972  & 4.9886  & 4.9894  &       & 4.9945  & 4.9995  & 4.9993  \bigstrut[t]\\
          & $\beta_2$ & -5.0186  & -5.0254  & -5.0274  &       & -5.0047  & -5.0051  & -5.0051  &       & -5.0021  & -5.0084  & -5.0082  \bigstrut[b]\\
    \hline
    \multirow{2}[2]{*}{S.D.} & $\beta_1$ & 0.0855  & 0.1022  & 0.1013  &       & 0.0624  & 0.0756  & 0.0749  &       & 0.0464  & 0.0561  & 0.0558  \bigstrut[t]\\
          & $\beta_2$ & 0.0715  & 0.0915  & 0.0915  &       & 0.0540  & 0.0701  & 0.0701  &       & 0.0453  & 0.0593  & 0.0595  \bigstrut[b]\\
    \hline
    \multirow{2}[2]{*}{MSE} & $\beta_1$ & 0.0147  & 0.0177  & 0.0178  &       & 0.0094  & 0.0116  & 0.0114  &       & 0.0036  & 0.0053  & 0.0055  \bigstrut[t]\\
          & $\beta_2$ & 0.0122  & 0.0167  & 0.0179  &       & 0.0073  & 0.0099  & 0.0101  &       & 0.0048  & 0.0081  & 0.0082  \bigstrut[b]\\
    \hline
    \hline
          &       &       &       &       &       &       & $df = 6$ &       &       &       &       &  \bigstrut\\
    \hline
    \multirow{2}[2]{*}{Mean} & $\beta_1$ & 5.0077  & 5.0118  & 5.0113  &       & 5.0035  & 5.0005  & 5.0000  &       & 5.0023  & 5.0026  & 5.0025  \bigstrut[t]\\
          & $\beta_2$ & -4.9980  & -4.9982  & -4.9987  &       & -5.0019  & -5.0040  & -5.0043  &       & -5.0026  & -5.0044  & -5.0042  \bigstrut[b]\\
    \hline
    \multirow{2}[2]{*}{S.D.} & $\beta_1$ & 0.0758  & 0.0843  & 0.0836  &       & 0.0530  & 0.0591  & 0.0584  &       & 0.0469  & 0.0530  & 0.0528  \bigstrut[t]\\
          & $\beta_2$ & 0.0730  & 0.0875  & 0.0871  &       & 0.0531  & 0.0643  & 0.0638  &       & 0.0402  & 0.0485  & 0.0483  \bigstrut[b]\\
    \hline
    \multirow{2}[2]{*}{MSE} & $\beta_1$ & 0.0108  & 0.0124  & 0.0123  &       & 0.0053  & 0.0063  & 0.0062  &       & 0.0046  & 0.0059  & 0.0059  \bigstrut[t]\\
          & $\beta_2$ & 0.0128  & 0.0161  & 0.0160  &       & 0.0067  & 0.0082  & 0.0082  &       & 0.0036  & 0.0047  & 0.0048  \bigstrut[b]\\
    \hline
    \hline
          &       &       &       &       &       &       & $df = \infty$ &       &       &       &       &  \bigstrut\\
    \hline
    \multirow{2}[2]{*}{Mean} & $\beta_1$ & 5.0003  & 5.0006  & 5.0005  &       & 5.0000  & 5.0046  & 5.0047  &       & 4.9941  & 4.9943  & 4.9944  \bigstrut[t]\\
          & $\beta_2$ & -4.9987  & -4.9942  & -4.9943  &       & -4.9946  & -4.9991  & -4.9991  &       & -4.9991  & -5.0015  & -5.0014  \bigstrut[b]\\
    \hline
    \multirow{2}[2]{*}{S.D.} & $\beta_1$ & 0.0695  & 0.0741  & 0.0738  &       & 0.0452  & 0.0478  & 0.0477  &       & 0.0389  & 0.0417  & 0.0416  \bigstrut[t]\\
          & $\beta_2$ & 0.0661  & 0.0763  & 0.0761  &       & 0.0438  & 0.0505  & 0.0505  &       & 0.0348  & 0.0398  & 0.0397  \bigstrut[b]\\
    \hline
    \multirow{2}[2]{*}{MSE} & $\beta_1$ & 0.0087  & 0.0097  & 0.0097  &       & 0.0042  & 0.0047  & 0.0047  &       & 0.0032  & 0.0036  & 0.0036  \bigstrut[t]\\
          & $\beta_2$ & 0.0093  & 0.0111  & 0.0111  &       & 0.0038  & 0.0045  & 0.0045  &       & 0.0027  & 0.0032  & 0.0032  \bigstrut[b]\\
    \hline
    \end{tabular}%
  \label{tab:tab: rem sim results t}%
\end{table}

The posterior standard deviations of our \textsc{dp-rem} are the smallest in all scenarios, indicating that its posteriors are more concentrated than the \textsc{dir-rem} and \textsc{nor-rem} ones. We also observe that for a given sample size the advantages of the \textsc{dp-rem} over the parametric \textsc{dir-rem} and \textsc{nor-rem} with respect to posterior standard deviations are the largest when $df = 2$, and become smaller as $df$ increases. This is the result of the tails of the t distributions mixed in the error distributions becoming less heavy. In addition, the \textsc{dir-rem} gives almost the same posterior standard deviations as the \textsc{nor-rem} when $df$ are 4, 6 and $\infty$. This demonstrates that the advantages of the \textsc{dp-rem} over the \textsc{dir-rem} and \textsc{nor-rem} fall at a slower rate than that of the \textsc{dir-rem} over the \textsc{nor-rem} when $df$ increases.

Our \textsc{dp-rem} dominates the two parametric estimators regarding \textsc{mse} under all circumstances. Similar to the posterior standard deviations, the advantages of the \textsc{dp-rem} over the \textsc{dir-rem} and \textsc{nor-rem} in terms of \textsc{mse} fall as the $df$ increases, as the tails of the mixed t distributions become less heavy. It should be noted that the \textsc{mse} of \textsc{dir-rem} and \textsc{nor-rem} are almost identical when $df = 4$. That is, the advantage of the \textsc{dir-rem} over the \textsc{nor-rem} has diminished. In contrast, the advantage of the \textsc{dp-rem} over the two parametric estimators is still present. This demonstrates that the parametric \textsc{dir-rem} and \textsc{nor-rem} are not as good as the semi-parametric \textsc{dp-rem} at identifying the heterogeneity in error distributions when the tails become less heavy.

\subsection{DP-REM Empirical Examples} \label{sec:DP-REM empirical}
As a demonstration of the \textsc{dp-rem} with real data, in this section we present the results regarding a model for wages of U.S. workers. The data are from \hyperlink{Cornwell and Rupert, 1988}{Cornwell and Rupert (1988)} with 595 individuals over a period of 7 years, from 1976 to 1982. In fact, it allows us to demonstrate the correlated \textsc{rem} in \eqref{CREM}, as outlined below. The model is given by
\begin{equation} \label{wage model}
\text{ln} Wage_{it} = \beta_1 Exp_{it} + \beta_2 Mar_{it} + \beta_3 Fem_{i}
+ \beta_4 Edu_{it} + v_i + \varepsilon_{it},
\end{equation}
where the dependent variable is the logarithms of wages, and the explanatory variables are experience in years ($Exp$), dummies for marriage status ($Mar$) and the individual being female ($Fem$), as well as the years of education ($Edu$). As there are strong reasons to suspect that the unobserved individual effect $v_i$ is correlated with the explanatory variables due to omitted variables such as ability and motivation, the correlated \textsc{rem} is selected. It specifies $v_i$ as
\begin{equation} \label{crem part wage}
v_i = \tilde{\beta_1} AExp_i + \tilde{\beta_2} AMar_i + u_i,
\end{equation}
where the sample averages of experience ($AExp$) and marriage status ($AMar$) of individual $i$ are included, as they are the two time variant variables in (\ref{wage model}). The posterior means, standard deviations and p-values of the parameters with the three estimators \textsc{dp-rem}, \textsc{dir-rem} and \textsc{nor-rem} are presented in Table \ref{tab:U.S. Individual Wage}. We also demonstrate their respective performances with a number of figures.

From Table \ref{tab:U.S. Individual Wage} we observe that the posterior means are similar for all parameters. Among the explanatory variables, experience and education are positively correlated with workers' wages, while being married and female negatively influences wages. The semi-parametric \textsc{dp-rem} gives the smallest posterior standard deviations among all three estimators. As the posterior modes of the numbers of clusters for $u_i$ and $\varepsilon_i$ are 2 and 6, respectively, heterogeneity is detected in both the individual effects' and the idiosyncratic errors' distributions. The posterior p-values of the \textsc{dp-rem} are also the smallest for all parameters. It should be noted that both parameters in \eqref{crem part wage} have posterior p-values less than 5\%, indicating that $v_i$ are correlated with the explanatory variables, providing support for the application of the correlated \textsc{rem}.

\begin{table}[htbp]
  \centering
  \scriptsize
  \setstretch{1.5}
  \caption{U.S. Individual Wage}
    \begin{tabular}{cccccccccccc}
    \toprule
    \toprule
          &       & Mean  &       &       &       & SD    &       &       &       & p-values     &  \\
    \midrule
    Parameter & DP    & Dir   & Nor   &       & DP    & Dir   & Nor   &       & DP    & Dir   & Nor \\
    \midrule
    $Exp$ & 0.0912  & 0.0968  & 0.0969  &       & 0.0007  & 0.0013  & 0.0013  &       & 0.0000  & 0.0000  & 0.0000  \\
    $Mar$ & -0.0194  & -0.0323  & -0.0324  &       & 0.0114  & 0.0197  & 0.0216  &       & 0.0880  & 0.0920  & 0.1520  \\
    $Fem$ & -0.3259  & -0.3160  & -0.3110  &       & 0.0592  & 0.0639  & 0.0777  &       & 0.0000  & 0.0000  & 0.0000  \\
    $Edu$ & 0.0771  & 0.0731  & 0.0738  &       & 0.0041  & 0.0048  & 0.0049  &       & 0.0000  & 0.0000  & 0.0000  \\
    $AExp$ & -0.0823  & -0.0885  & -0.0885  &       & 0.0015  & 0.0018  & 0.0020  &       & 0.0000  & 0.0000  & 0.0000  \\
    $AMar$ & 0.1578  & 0.1796  & 0.1850  &       & 0.0521  & 0.0602  & 0.0697  &       & 0.0000  & 0.0020  & 0.0080  \\
    \bottomrule
    \end{tabular}%
  \label{tab:U.S. Individual Wage}%
\end{table}

\begin{figure}[h!]
 \centering
 \caption{95\% \textsc{hpdi}'s of Parameters in the U.S. Wage Model} \label{fig: CI-REM}
 \includegraphics[width = 0.9\linewidth]{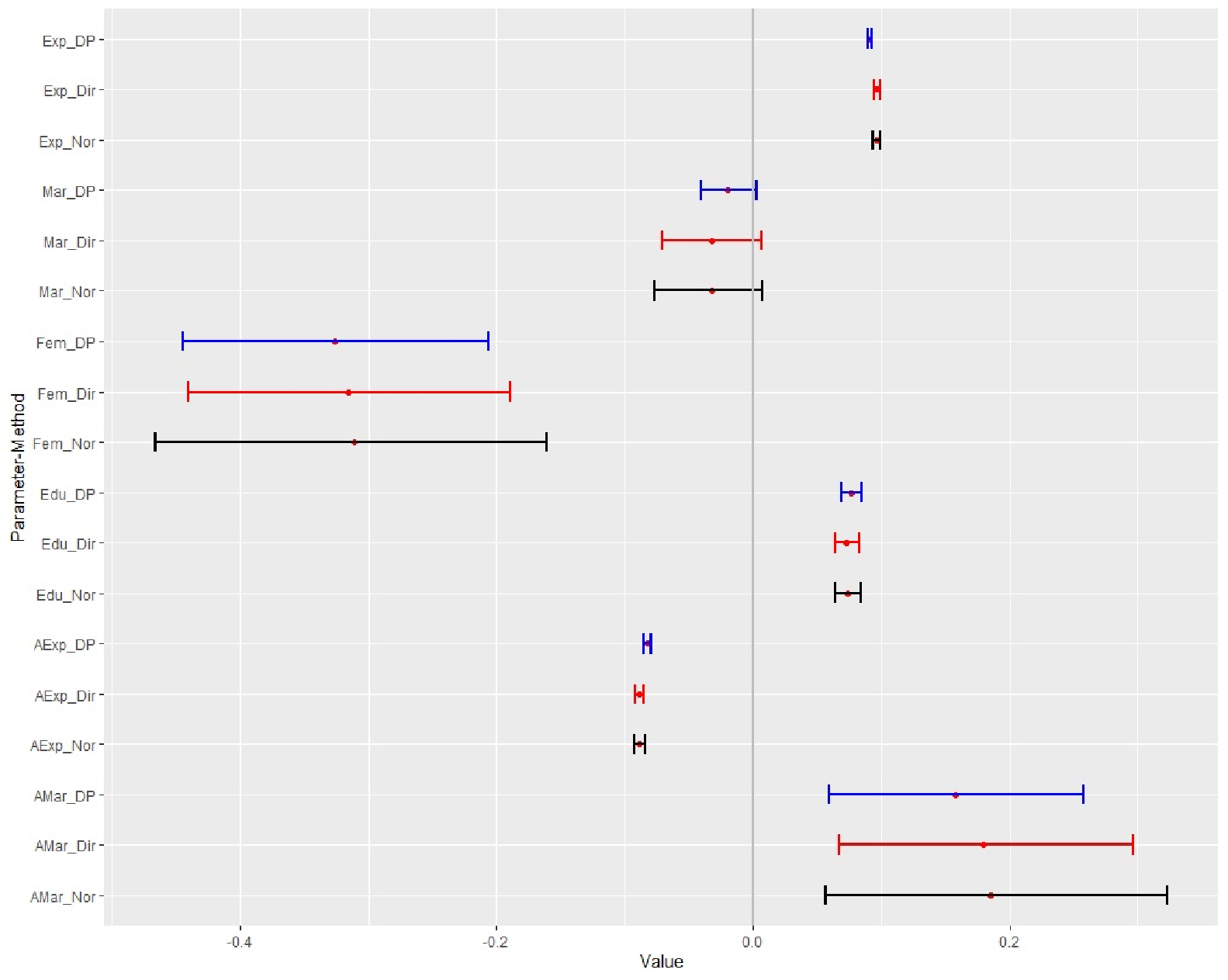}
 \begin{minipage}{0.8\linewidth}
    \footnotesize
    \setstretch{1.5}
    Note: For the parameter of each variable (e.g. ``Edu'' for education), we present the 95\% HPDI's of all three estimators: \textsc{dp-rem}, \textsc{dir-rem} and \textsc{nor-rem}. The three estimators are also allocated with the colours blue, red and black in the same order.
 \end{minipage}
\end{figure}

In Figure \ref{fig: CI-REM} we present the 95\% \textsc{hpdi}'s of the three estimators for all regression parameters in \eqref{crem part wage}. One can observe that the semi-parametric \textsc{dp-rem} yields the narrowest 95\%  \textsc{hpdi}'s for all parameters, followed by the \textsc{dir-rem} with a parametric mixture. The parametric \textsc{nor-rem} has the widest \textsc{hpdi}'s for all the parameters, though for some parameters its \textsc{hpdi}'s are rather similar to those of \textsc{dir-rem}, e.g. with the parameter $\beta_4$ for education.

\begin{figure}[h!]
\begin{center}
\caption{Histograms of the Parameter for Education} \label{fig:Hist DP-REM}
\begin{subfigure}{.3\textwidth}
  \centering
  \includegraphics[width=.99\linewidth]{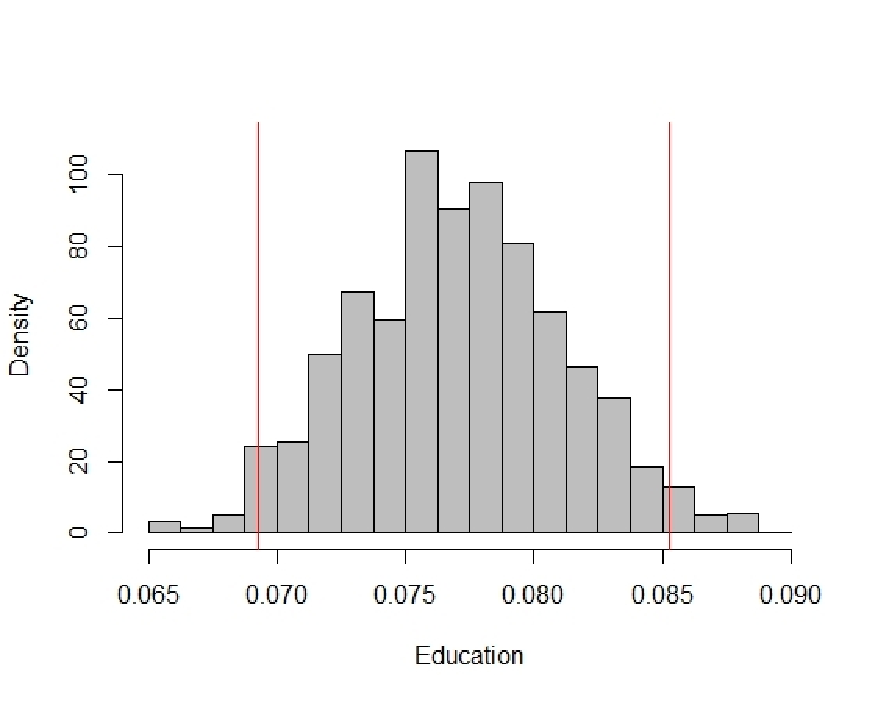}
  \caption{\textsc{dp-rem}}
  \label{fig2:sub-first}
\end{subfigure}
\begin{subfigure}{.3\textwidth}
  \centering
  \includegraphics[width=.99\linewidth]{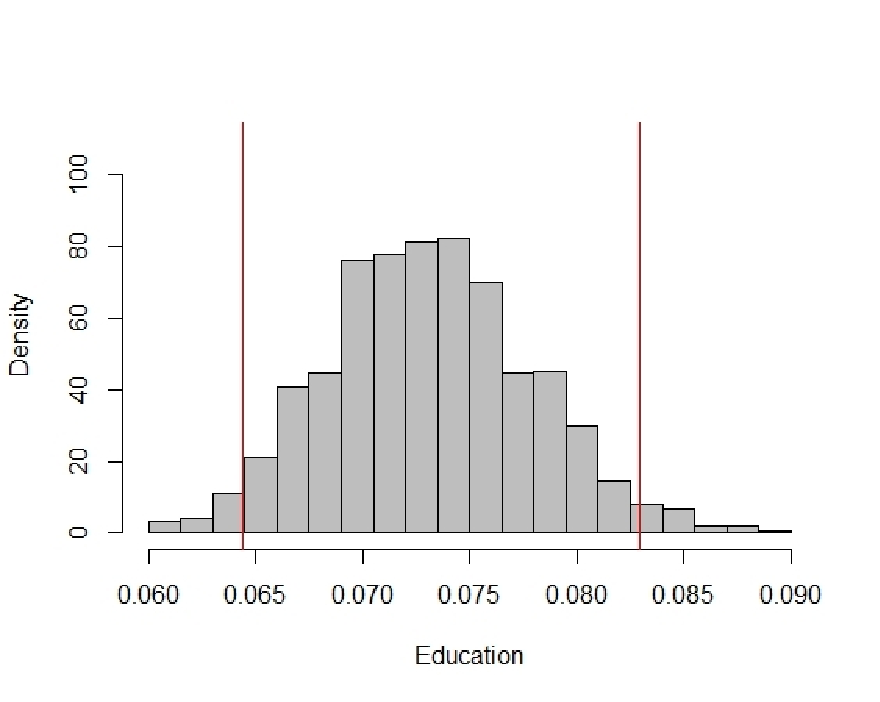}
  \caption{\textsc{dir-rem}}
  \label{fig2:sub-second}
\end{subfigure}
\begin{subfigure}{.3\textwidth}
  \centering
  \includegraphics[width=.99\linewidth]{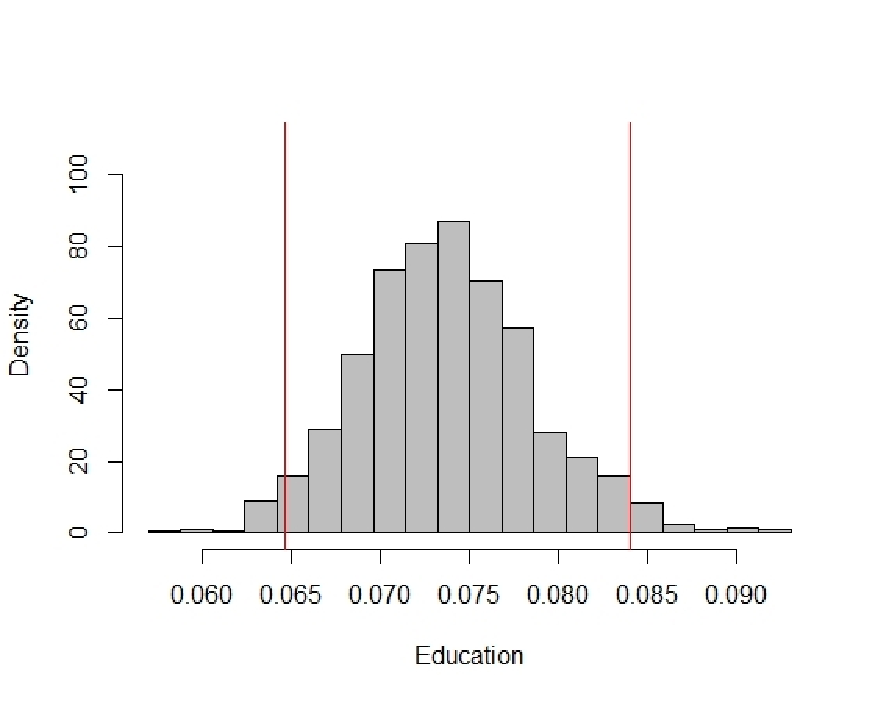}
  \caption{\textsc{nor-rem}}
  \label{fig2:sub-third}
\end{subfigure}
\end{center}
\begin{center}
\begin{minipage}{0.8\linewidth}
  \footnotesize
  \setstretch{1.5}
  Note: The three panels (\ref{fig2:sub-first}), (\ref{fig2:sub-second}) and (\ref{fig2:sub-third}) are for the \textsc{dp-rem}, \textsc{dir-rem} and \textsc{nor-rem}, respectively. The interval is $(0.0692, 0.0852)$ for \textsc{dp-rem}, $(0.0643, 0.0829)$ for \textsc{dir-rem}, and $(0.0646, 0.0840)$ for \textsc{nor-rem}. The lower and upper bounds of the 95\% \textsc{hpdi}'s are marked by (red) vertical lines.
\end{minipage}
\end{center}
\end{figure}

Figure \ref{fig:Hist DP-REM} presents the histograms of the posterior draws for the education parameter ($\beta_4$). Though all three intervals exclude zero, the \textsc{dp-rem} one is the shortest, as a result of the efficiency gain from exploring the heterogeneity in the error distributions.

\begin{figure}[h!]
\begin{center}
\caption{Predictive Densities of Individual Effects and Residuals} \label{fig:Pred Den DP-REM}
\begin{subfigure}{0.45\textwidth}
  \centering
  \includegraphics[width=.99\linewidth]{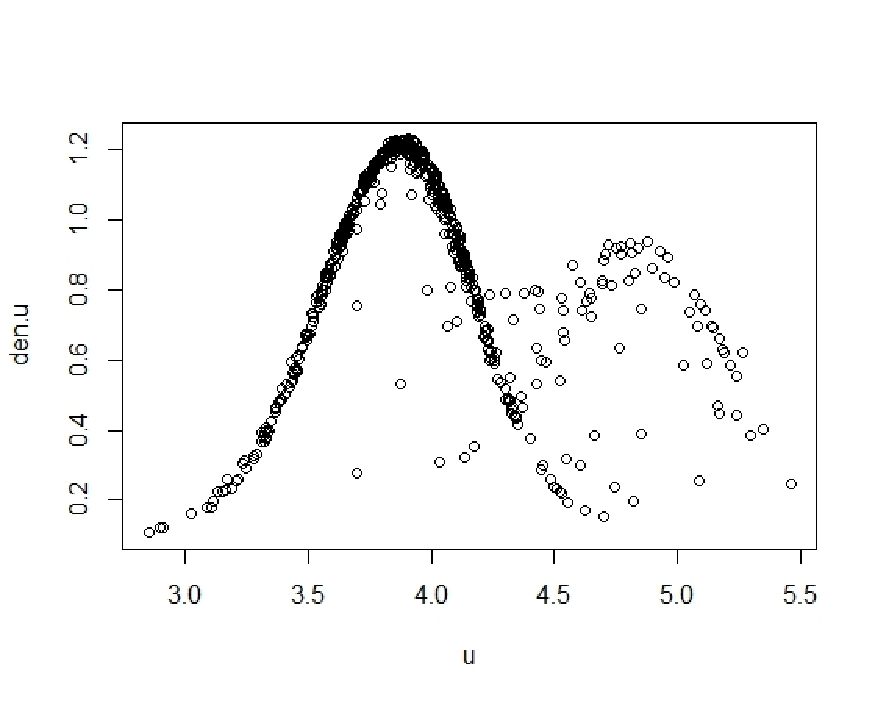}
  \caption{u, \textsc{dp-rem}}
  \label{fig: 4-a}
\end{subfigure}
\begin{subfigure}{0.45\textwidth}
  \centering
  \includegraphics[width=.99\linewidth]{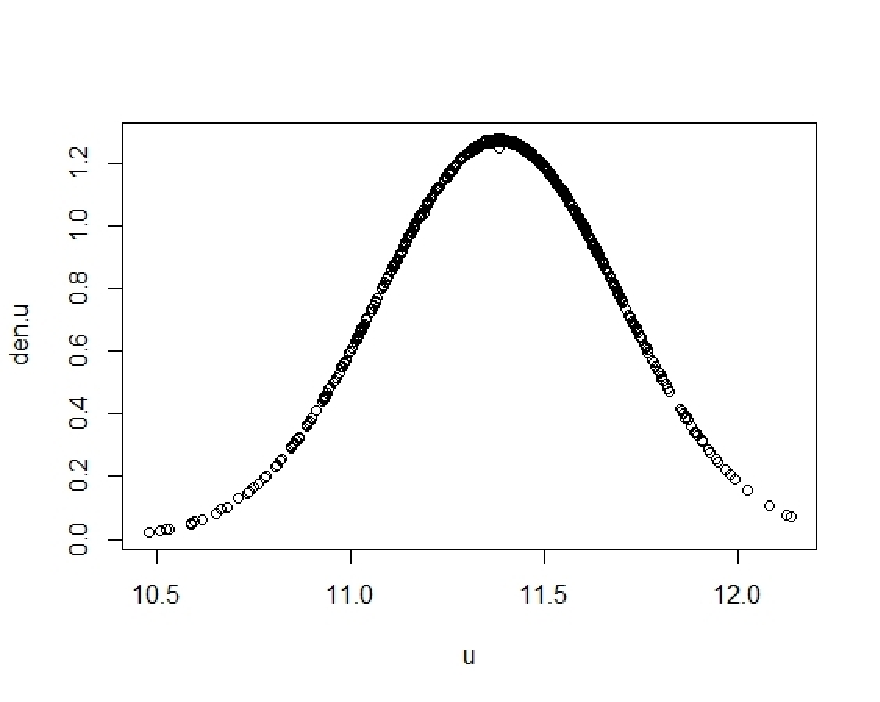}
  \caption{u, \textsc{nor-rem}}
  \label{fig: 4-b}
\end{subfigure}
\\
\begin{subfigure}{0.45\textwidth}
  \centering
  \includegraphics[width=.99\linewidth]{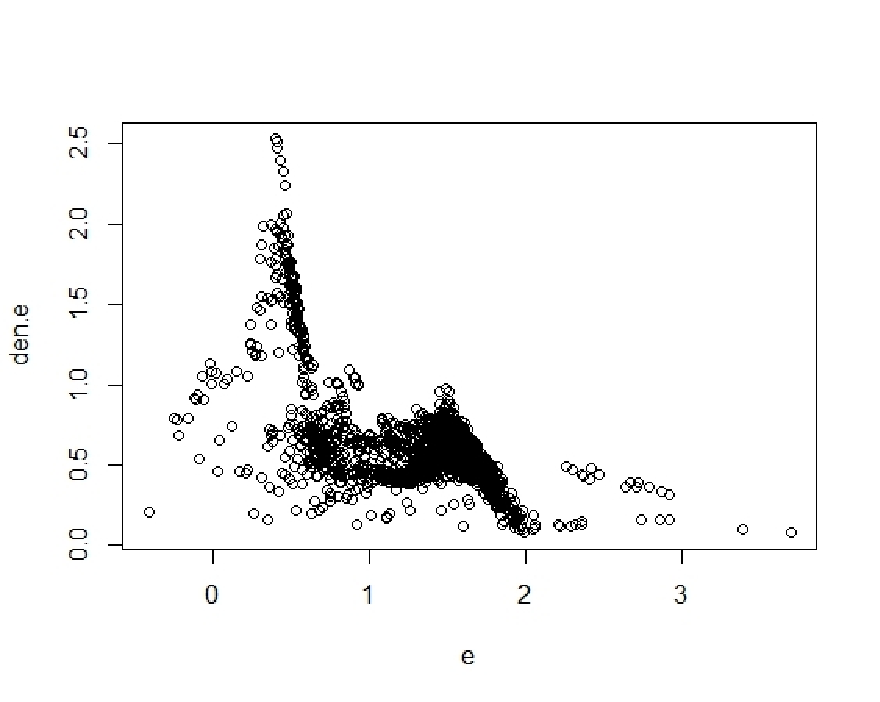}
  \caption{e, \textsc{dp-rem}}
  \label{fig: 4-c}
\end{subfigure}
\begin{subfigure}{0.45\textwidth}
  \centering
  \includegraphics[width=.99\linewidth]{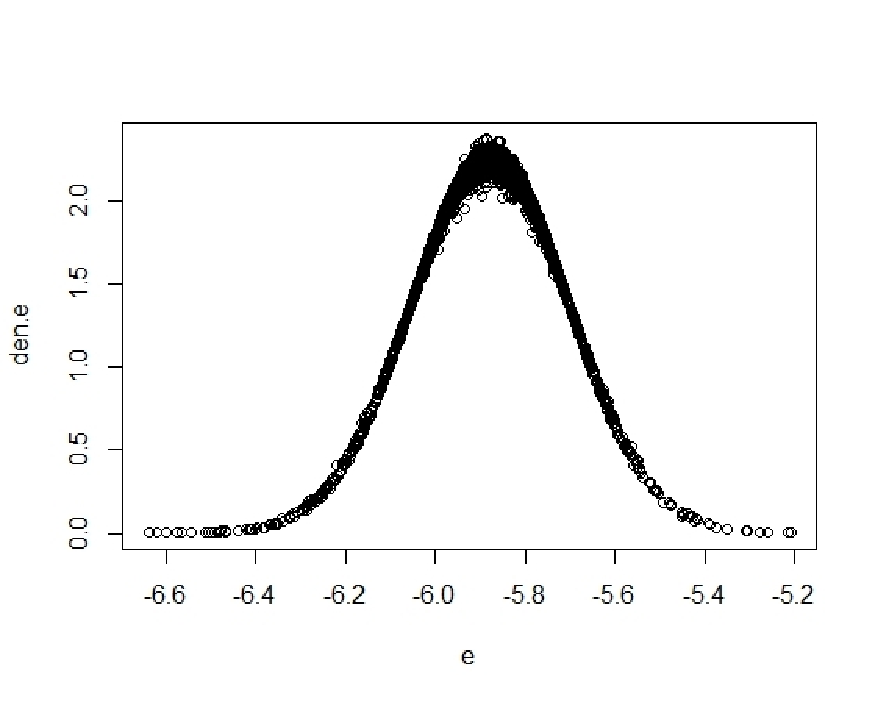}
  \caption{e, \textsc{nor-rem}}
  \label{fig: 4-d}
\end{subfigure}
\end{center}
\begin{center}
\begin{minipage}{0.8\linewidth}
	\footnotesize
    \setstretch{1.5}
	Note: The predictive densities of the individual effects are in panels (\ref{fig: 4-a}) (\textsc{dp-rem}) and (\ref{fig: 4-b}) (\textsc{nor-rem}). The predictive densities of the residuals are in panel (\ref{fig: 4-c}) (\textsc{dp-rem}) and (\ref{fig: 4-d}) (\textsc{nor-rem}).
\end{minipage}
\end{center}
\end{figure}

Figure \ref{fig:Pred Den DP-REM} presents the predictive densities of the fitted individual effects and residuals obtained by the semi-parametric \textsc{dp-rem} and the parametric \textsc{nor-rem}, respectively. One can see that both the fitted individual effects and the residuals have multi-modal distributions with our \textsc{dp-rem}, which is not captured by the \textsc{nor-rem}. Thus, the normal distribution assumption could cause potential losses in efficiency.

\section{Conclusion} \label{sec:Conclusion}
In this paper we address the potential violation of the assumptions made by parametric Bayesian \textsc{gls} estimators that the errors are homogeneous regarding their distributions. Such assumptions are likely to be problematic in reality particularly when micro data are used, as the features of individuals or households are likely to lead to observations having different distributions. We present a semi-parametric Bayesian \textsc{gls} where the error distribution is a non-parametric mixture of normal distributions by introducing a Dirichlet process prior on the distributional parameters of the errors. The number of normal components is decided jointly by the data and the prior in such a mixture, which is able to cover a large variety of distributions. The errors are grouped by the $\mathcal{DP}$ prior, with those in the same group having the same distributional parameters and thus the same distribution. Two specific cases of the semi-parametric Bayesian \textsc{gls} are then introduced, which are the \textsc{sur} for equation systems and the \textsc{rem} for panel data.

Our \textsc{dp-sur} and \textsc{dp-rem} methods are demonstrated with a series of simulation experiments consisting of two scenarios, where the errors have a log-normal distribution and a mixture of t distributions that are bi-modal and asymmetric, respectively. When the errors have a log-normal distribution, which is fat tailed, our semi-parametric \textsc{gls} estimator gives smaller posterior standard deviations, as well as smaller mean squared errors in all settings. Such advantages over the parametric estimators are greater when the variances of the errors are larger, leading to  heavier tails of the error distributions. When the errors are bi-modal as a mixture of t distributions, our semi-parametric \textsc{gls} estimators also out-perform the parametric estimators with respect to posterior standard errors and mean squared errors in all scenarios. Such advantages are larger when the degrees of freedom of the t mixture components are smaller, i.e., the tails of the mixture are heavier.

We apply our \textsc{dp-sur} method to the demands for production factors with the generalized Leontief cost function using a dataset of the U.S. banking industry. Heterogeneity is detected in the sample by our semi-parametric estimator. The \textsc{dp-sur} posterior standard deviations are smaller than the \textsc{dir-sur} ones using a parametric mixture of normal distributions, as well as the \textsc{nor-sur} ones for all the demand elasticities. In addition, the posterior p-values of the \textsc{dp-sur} are also smaller for all the parameters.

Our \textsc{dp-rem} is also applied to a study U.S. workers' wages, where there is a strong reason to suspect that the unobserved individual effects are correlated with the explanatory variables due to omitted variables describing individual features such as abilities. The correlated \textsc{rem} is then estimated. The \textsc{dp-rem} detects heterogeneity in the distributions of both the individual effects and the idiosyncratic errors with this sample. Our \textsc{dp-rem} obtains smaller posterior standard deviations and p-values than the parametric \textsc{dir-rem} and \textsc{nor-rem} for all parameters.

\pagebreak
\section*{Acknowledgement}
We are grateful to the useful comments received from Debopam Bhattacharya, Xiaohong Chen, Gernot Doppelhofer, Oliver Linton and Justin Tobias.

\pagebreak

\end{document}